\documentclass[longauth]{aa}

\usepackage{graphicx}
\usepackage[dvipsnames]{xcolor}
\usepackage{txfonts}
\usepackage[colorlinks=true, citecolor=blue]{hyperref}

\usepackage{multirow}
\usepackage{booktabs}
\usepackage{lscape}
\usepackage{placeins}
\usepackage{caption}

\newcommand{\hb}{H$\beta$}
\newcommand{\ha}{H$\alpha$}

\newcommand{\neiii}{[Ne\,{\footnotesize III}]}
\newcommand{\oiii}{[O\,{\footnotesize III}]}
\newcommand{\oii}{[O\,{\footnotesize II}]}
\newcommand{\nii}{[N\,{\footnotesize II}]}

\newcommand{\kms}{km s$^{-1}$}
\newcommand{\myr}{$M_{\odot}$ yr$^{-1}$}
\newcommand{\msun}{$M_{\odot}$}

\newcommand{\mstar}{$M_{\star}$}

\newcommand{\jwst}{JWST}
\newcommand{\hst}{HST}

\newcommand{\sigmastar}{$\sigma_\star$}
\newcommand{\mgii}{Mg\,{\sc ii}} 
\newcommand{\mgi}{Mg\,{\sc i}}
\newcommand{\feii}{Fe\,{\sc ii}}
\newcommand{\mout}{$M_{\rm out}$}
\newcommand{\mdotout}{$\dot{M}_{\rm out}$}

\newcommand{\snhope}{NS\_274}
\newcommand{\rubies}{RUBIES-UDS-QG-z7}

\begin{document} 

   \title{Gas outflows in two recently quenched galaxies at $z=4$ and $7$}

  \titlerunning{Gas outflows in two recently quenched galaxies at $z=4$ and $7$}
  \authorrunning{F. Valentino et al.}
  
   \author{F. Valentino\inst{1,2,3} \and  
           K. E. Heintz\inst{1,4,5} \and
           G. Brammer\inst{1,4} \and 
           K. Ito\inst{1,2,6} \and 
           V. Kokorev\inst{7} \and
           K. E. Whitaker\inst{8, 1} \and
           A. Gallazzi\inst{9} \and
           A. de Graaff\inst{10} \and
           A. Weibel\inst{5} \and
           B. L. Frye\inst{11} \and
           P. S. Kamieneski\inst{12} \and
           S. Jin\inst{1,2} \and
           D. Ceverino\inst{13,14} \and
           A. Faisst\inst{15} \and
           M. Farcy\inst{16} \and
           S. Fujimoto\inst{7} \and
           S. Gillman\inst{1,2} \and
           R. Gottumukkala\inst{1,4} \and
           M. Hamadouche\inst{8} \and
           K. C. Harrington\inst{17,18,19,20} \and      
           M. Hirschmann\inst{16} \and
           C. K. Jespersen\inst{21} \and
           T. Kakimoto\inst{22,23} \and
           M. Kubo\inst{24,25} \and
           C. d. P. Lagos\inst{26,1} \and
           M. Lee\inst{1,2} \and
           G. E. Magdis \inst{1,2} \and
           A. W. S. Man\inst{27} \and
           M. Onodera\inst{22,23,28} \and
           F. Rizzo\inst{29} \and
           R. Shimakawa\inst{30} \and
           D. J. Setton\inst{31} \and
           M. Tanaka\inst{22,23} \and
           S. Toft\inst{1,4} \and
           P.-F. Wu\inst{32} \and
           P. Zhu\inst{1,2}}
    \institute{
    Cosmic Dawn Center (DAWN), Denmark
    \and DTU Space, Technical University of Denmark, Elektrovej 327, DK-2800 Kgs. Lyngby, Denmark
    \and European Southern Observatory, Karl-Schwarzschild-Str. 2, 85748 Garching, Germany
    \and Niels Bohr Institute, University of Copenhagen, Jagtvej 128, 2200, Copenhagen N, Denmark
    \and Department of Astronomy, University of Geneva, Chemin Pegasi 51, 1290 Versoix, Switzerland
    \and Department of Astronomy, School of Science, The University of Tokyo, 7-3-1, Hongo, Bunkyo-ku, Tokyo, 113-0033, Japan 
    \and Department of Astronomy, The University of Texas at Austin, Austin, TX 78712, USA
    \and Department of Astronomy, University of Massachusetts, Amherst, MA 01003, USA
    \and INAF - Osservatorio Astrofisico di Arcetri, Largo Enrico Fermi 5, 50125 Firenze, Italy
    \and Max-Planck-Institut f\"{u}r Astronomie, K\"{o}nigstuhl 17, D-69117, Heidelberg, Germany
    \and Department of Astronomy/Steward Observatory, University of Arizona, 933 N Cherry Ave., Tucson, AZ 85721-0009, USA
    \and School of Earth and Space Exploration, Arizona State University, PO Box 876004, Tempe, AZ 85287-6004, USA
    \and Departamento de Fisica Teorica, Modulo 8, Facultad de Ciencias, Universidad Autonoma de Madrid, 28049 Madrid, Spain
    \and CIAFF, Facultad de Ciencias, Universidad Autonoma de Madrid, 28049 Madrid, Spain
    \and Caltech/IPAC, MS 314-6, 1200 E. California Blvd. Pasadena, CA 91125, USA
    \and Institute of Physics, Laboratory for Galaxy Evolution, EPFL, Observatoire de Sauverny, Chemin Pegasi 51, 1290 Versoix, Switzerland 
    \and Joint ALMA Observatory, Alonso de C{\'o}rdova 3107, Vitacura, Casilla 19001, Santiago de Chile, Chile
    \and National Astronomical Observatory of Japan, Los Abedules 3085 Oficina 701, Vitacura 763 0414, Santiago, Chile
    \and European Southern Observatory, Alonso de C{\'o}rdova 3107, Vitacura, Casilla 19001, Santiago de Chile, Chile
    \and Universidad Diego Portales, Av. Ejercito Santiago de Chile, Chile
    \and Department of Astrophysical Sciences, Princeton University, Princeton, NJ 08544, USA
    \and Department of Astronomical Science, The Graduate University for Advanced Studies, SOKENDAI, 2-21-1 Osawa, Mitaka, Tokyo 181-8588, Japan
    \and National Astronomical Observatory of Japan, 2-21-1 Osawa, Mitaka, Tokyo 181-8588, Japan
    \and School of Science, Kwansei Gakuin University, 2-1, Gakuen, Sanda, Hyogo 669-1337, Japan
    \and Astronomical Institute, Tohoku University, Aoba-ku, Sendai 980-8578, Japan
    \and International Centre for Radio Astronomy Research (ICRAR), M468, University of Western Australia, 35 Stirling Hwy, Crawley, WA 6009, Australia
    \and Department of Physics \& Astronomy, University of British Columbia, 6224 Agricultural Road, Vancouver BC, V6T 1Z1, Canada
    \and Subaru Telescope, National Astronomical Observatory of Japan, National Institutes of Natural Sciences (NINS), 650 North A'ohoku Place, Hilo, HI 96720, USA
    \and Kapteyn Astronomical Institute, University of Groningen, Landleven 12, 9747 AD, Groningen, The Netherlands 
    \and Waseda Institute for Advanced Study (WIAS), Waseda University, 1-21-1 Nishi-Waseda, Shinjuku, Tokyo 169-0051, Japana
    \and Center for Interdisciplinary Exploration and Research in Astrophysics (CIERA), Northwestern University,1800 Sherman Ave, Evanston, IL 60201, USA
    \and Graduate Institute of Astrophysics, National Taiwan University, Taipei 10617, Taiwan
}       

   \date{Received --; accepted --}

  \abstract{Outflows are a key element in the baryon cycle of galaxies, impacting their evolution by extracting gas, momentum, and energy and then injecting them into the surrounding medium. 
  The properties of gas outflows provide a fundamental test for our models of how galaxies transition from a phase of active star formation to quiescence. Here we report the detection of outflowing gas signatures in two recently quenched, massive ($M_\star \sim 10^{10.2}\,M_\odot$) galaxies at $z=4.106$ (\snhope) and $z=7.276$ (\rubies) observed at rest-frame ultraviolet to near-infrared wavelengths with JWST/NIRSpec. The outflows are traced by blueshifted magnesium (MgII) absorption lines, and in the case of the $z=4.1 $ system, also by  iron (FeII) and sodium (NaI) features. Together, these transitions broadly trace the chemically enriched neutral phase of the gaseous medium. The rest-frame optical spectra of the two sources are similar to those of local post-starburst galaxies, showing deep Balmer stellar features, a relatively low $D_n4000$ index, and minimal ongoing star formation on 10 Myr timescales, as traced by the lack of bright nebular and recombination emission lines.\ This also suggests the absence of strong and radiatively efficient active galactic nucleus activity. The galaxies' star formation histories are consistent with a recent and abrupt quenching, prior to which the average star formation rate was $\sim15$ \myr\ over the last 100 Myr of their lives. In the case of \snhope, dedicated millimeter observations allowed us to also strongly constrain the dust obscured star formation rate to $<12$\myr, unambiguously confirming its quiescence. Under simple geometrical assumptions, we derive mass loading factors $\eta=\dot{M}_{\rm out}/\mathrm{SFR_{100\,Myr}}\lesssim1$ and $>10$ for the $z=4.1$ and $z=7.3$ systems, respectively, and a similarly pronounced difference in the energy carried by the outflows.
  Supernova feedback could account for the mass and energy of the outflow in \snhope. However, the low mass loading factor and average gas velocity ($\sim180$ \kms, which is lower than the stellar velocity dispersion) suggest that the observed outflow is likely not the primary factor behind the quenching of \snhope, but it might represent a relic of the star formation process winding down. Star-formation-related processes seem to also be insufficient to explain the extreme mass outflow rate of \rubies, which would require an additional ejective mechanism such as an undetected active galactic nucleus. Finally, the average outflow velocities per unit stellar mass, star formation rate, and surface density of star formation rate are consistent with those of lower-redshift post-starburst galaxies, suggesting that outflows in rapidly quenched galaxies might occur similarly across cosmic time. Our findings hint at the existence of a rich tapestry of galaxy quenching pathways at high redshifts, and they highlight the importance of using large spectroscopic samples that map different spectral features to account for the different timescales on which different mechanisms contribute to this process.}  
  
   \maketitle

\section{Introduction}

The physical mechanisms responsible for the suppression of star formation in galaxies remain a major topic of debate in modern astrophysics, especially after the spectroscopic confirmation of massive (stellar masses of $M_\star\sim10^{10-11}$ \msun) ``quenched'' systems in the first few hundred million years ($z\sim3-7$) of our Universe's history (\citealt{glazebrook_2017, schreiber_2018c, tanaka_2019, valentino_2020a, forrest_2020a, forrest_2020b, forrest_2022,  carnall_2023b, carnall_2024, nanayakkara_2024, glazebrook_2024, urbano-stawinski_2024, tanaka_2024, kakimoto_2024, setton_2024, onoue_2024, kokorev_2024, antwi-danso_2025,  deGraaff_2025, turner_2025, wu_2025, weibel_2025, baker_2025} among others). Models and simulations have to invoke powerful winds and gas ejection powered by active galactic nuclei (AGNs) in order to reproduce stellar masses and stellar mass functions at $z=0$, with a lesser role played by star formation feedback, especially at the highest masses (see the reviews by \citealt{somerville-dave_2015, naab-ostriker_2017}). However, despite the huge effort to refine the feedback recipes implemented in models, the number density and stellar masses of quenched galaxies at $z>3$ represent a challenge in most, if not all, state-of-the-art cosmological simulations \citep[e.g.,][]{schreiber_2018c, merlin_2019, valentino_2020a, valentino_2023, delucia_2024, lagos_2025, xie_2024, baker_2025}.

\begin{figure}
    \centering
    \includegraphics[width=\columnwidth]{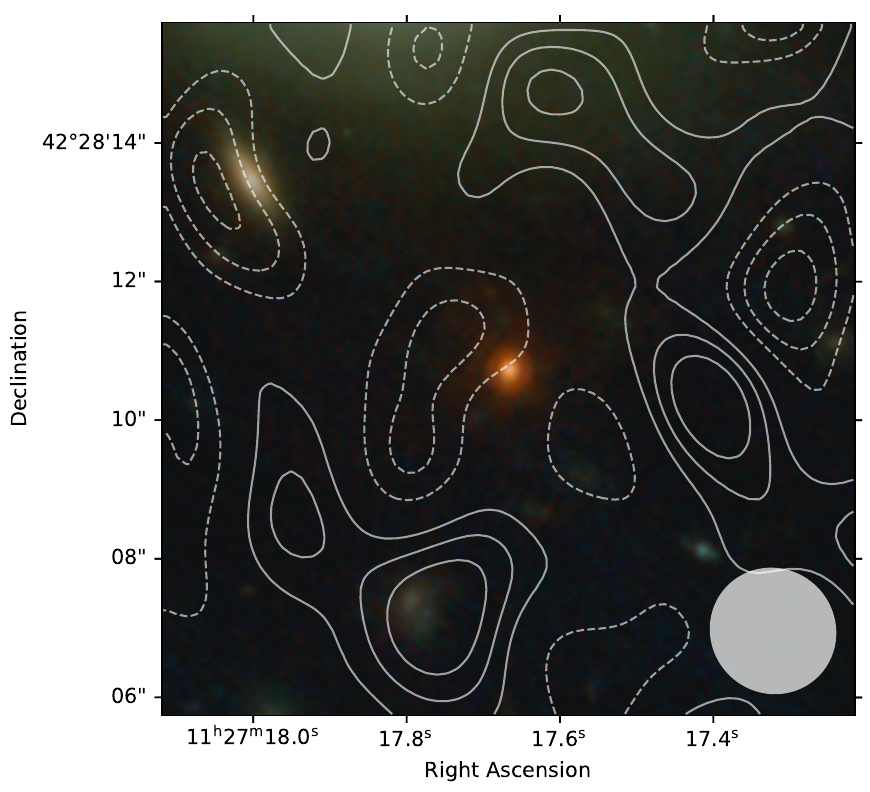}
    \caption{RGB color image of \snhope\ at $z=4.1061$ overlaid with NOEMA $1.3$ mm contours. The 10"$\times$10" background image is a color composite of six available wide-band NIRCam filters. The dashed and solid white contours indicate the $[-3\sigma,-2\sigma,-1\sigma,1\sigma,2\sigma,3\sigma]$ dust continuum emission at a $260\,\mu m$ rest frame wavelength, determined with NOEMA observations. The elliptical patch indicates the size of the beam.}
    \label{fig:rgb_noema}
\end{figure}

The exact details of the physical models of AGN and stellar feedback differ from simulation to simulation, but outflows remain a staple mechanism to quench galaxies in most available models \citep{kurinchi-vendhan_2024,lagos_2025}. Different  recipes and implementations might result in equally robust stellar mass functions in the local Universe, but they predict different outflow properties -- which thus become a powerful tool for gaining insight into the physics governing galaxies. Observationally, multiphase gas flowing outward from star-forming galaxies or systems harboring actively growing supermassive black holes is well documented, especially up to cosmic noon, based on the detection of both emission and absorption lines across the whole electromagnetic spectrum \citep[e.g.,][]{forster-schreiber_2020,veilleux_2020}. Focusing on the colder and denser outflow phases, the detection of low ionization absorption features in the rest-frame ultraviolet (UV) spectrum of galaxies has been a chief probe of their properties \citep[e.g.,][]{frye_2002, shapley_2003, rupke_2005, veilleux_2005, weiner_2009, rubin_2014, veilleux_2020, xu_2022, mingozzi_2022}. This allowed the outflow properties to be mapped  across a large dynamic range of host galaxy properties and various scaling relations to be established \citep[e.g., the average and maximum outflow velocity vs. stellar mass, star formation rate, and surface densities; see][and references therein]{davis_2023}. 
\begin{figure*}
    \centering
    \includegraphics[width=\textwidth]{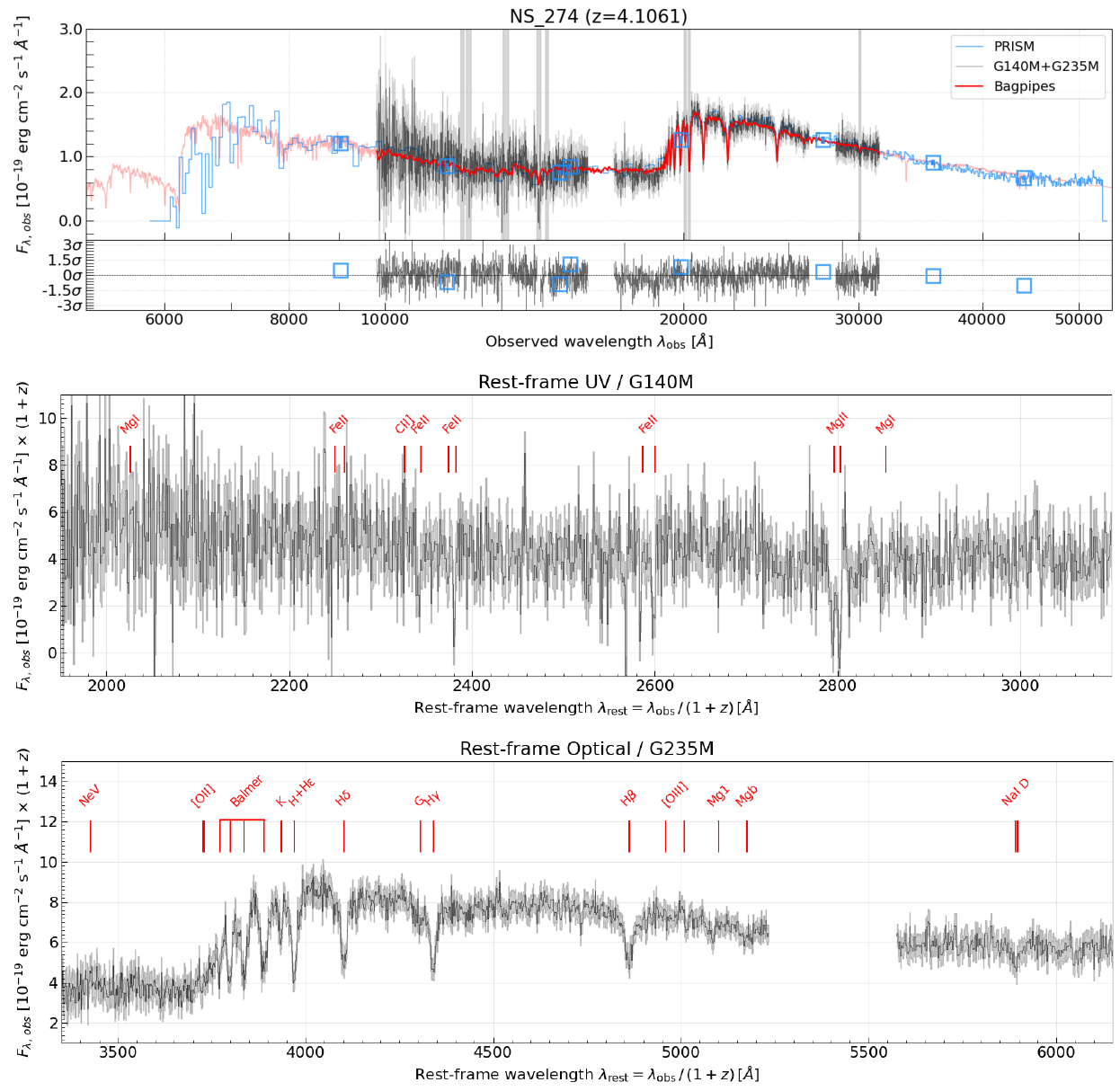}
    \caption{Spectra, photometry, and SED modeling of \snhope\ at $z=4.1061\pm0.0003$. \textit{Top:} Photometry (blue squares), JWST/NIRSpec medium resolution spectra (G140M+G235M, black line), and their uncertainties (gray line) that we jointly modeled with \textsc{Bagpipes} (red line) in the observed frame. The normalized residuals of the modeling are shown in the lower inset. For reference, we also show the PRISM spectrum (blue). 
    \textit{Center:} Rest-frame UV spectrum covered by the G140M grating. \textit{Bottom:} Rest-frame optical spectrum covered by the G235M grating. We mark in red the expected location of emission and absorption lines at the derived systemic redshift, $z_{\rm spec}=4.1061$, as labeled. The flux densities in the bottom panels have been rescaled by $(1+z)$ to conserve the energy.}
    \label{fig:sed_snhope}
\end{figure*}

In massive systems, the mixture of intense star formation and the simultaneous growth of central black holes makes the outflow phenomenon more complicated to decipher. Moreover, studying the population of recently quenched systems that likely experienced the effects of both supernova (SN) and AGN-powered feedback in their recent past, and are on their way to quiescence, has been a particularly difficult task to undertake given their relative rarity after cosmic noon and faintness at UV wavelengths. In general, most observational studies ascertained the presence of outflowing gas in recently quenched systems, both in the presence and absence of ongoing AGN activity \citep[e.g.,][]{tremonti_2007, coil_2011, maltby_2019, baron_2020, baron_2022, man_2021, davis_2023, taylor_2024, sun_2024}.

At higher redshifts, when approaching the main epoch of quenching of the most massive systems, galaxies showing typical ``post-starburst'' signatures, such as deep Balmer absorption lines, shallow $D_n4000$ breaks, and weak emission lines, naturally become more common \citep{d'eugenio_2020, forrest_2020b}. This makes the study of this population more appealing and urgent -- and, thanks to JWST, feasible. Interestingly, AGNs seem to be common in recently quenched galaxies at $z=2-3$ and above. Extreme optical line ratios, broad emission features, and mid-infrared detections are routinely reported \citep[e.g.,][]{carnall_2023b, d'eugenio_2024, belli_2024, nanayakkara_2024, kokorev_2024, onoue_2024, ji_2024}. Ionized outflows on their own \citep[e.g.,][]{kubo_2022} are likely not sufficient to quench star formation, but signatures of massive amounts of neutral gas leaving quenched galaxies at $z=2-3$ have recently been reported in the literature \citep{d'eugenio_2024, belli_2024, davies_2024}. The simultaneous presence of high-velocity ($\sim1000$ \kms) neutral outflows and AGN signatures has been interpreted as the smoking gun of a causal connection between the activity of supermassive black holes and quenching. However, the contribution to the mass and energy budget of cold outflows due to the last episode of star formation or its residuals is yet to be fully understood \citep[see, e.g.,][for an example of a starburst-driven outflow in a massive galaxy]{rupke_2019}. This is especially true in the presence of outflows less extreme in terms of velocity and mass.\\

In this work we studied two distant and recently quenched galaxies that show signatures of outflowing gas: \snhope\ at $z=4.106$ and \rubies\ at $z=7.276$. Their redshifts were first confirmed with JWST/NIRSpec observations in \cite{frye_2024} and \cite{weibel_2025}, respectively. These two sources stand out as the most distant quenched galaxies with high S/N coverage of the rest-frame UV and optical spectra at medium and high spectral resolution in our archival search and dedicated survey with JWST/NIRSpec (the ``DeepDive'' project; Ito et al. in prep.). They thus allowed us to explore a new redshift regime; future work on  population studies at lower redshifts ($z\sim2-3$) will follow. In particular, in this work we confirm and refine the redshift of \rubies\ thanks to newly acquired data with a G235M/F170LP grating and filter combination, and a custom reduction pipeline that is able to recover the whole wavelength coverage allowed by the long-pass filter and position on the detector (Brammer et al. in prep.). Here we mainly focus on the iron (Fe) and magnesium (Mg) absorption features in the rest-frame UV and their blueshifts, which indicate the presence of cold outflowing gas. In addition, we independently reanalyzed the properties of \snhope, which are presented in \cite{wu_2025}, and incorporated information from the sodium doublet NaI D at rest-frame optical wavelengths. The spectra and photometry analyzed in this work are available online\footnote{\href{https://doi.org/10.5281/zenodo.15518189}{10.5281/zenodo.15518189}}.\\

Throughout this work, we make use of the AB system to report magnitudes \citep{oke-gunn_1983}. We adopt a $\Lambda$ cold dark matter cosmology with $\Omega_{\rm m} = 0.3$, $\Omega_{\Lambda} = 0.7$, and $H_0 = 70\,\mathrm{km\,s^{-1}\,Mpc^{-1}}$.

\begin{figure*}
    \centering
   \includegraphics[width=\textwidth]{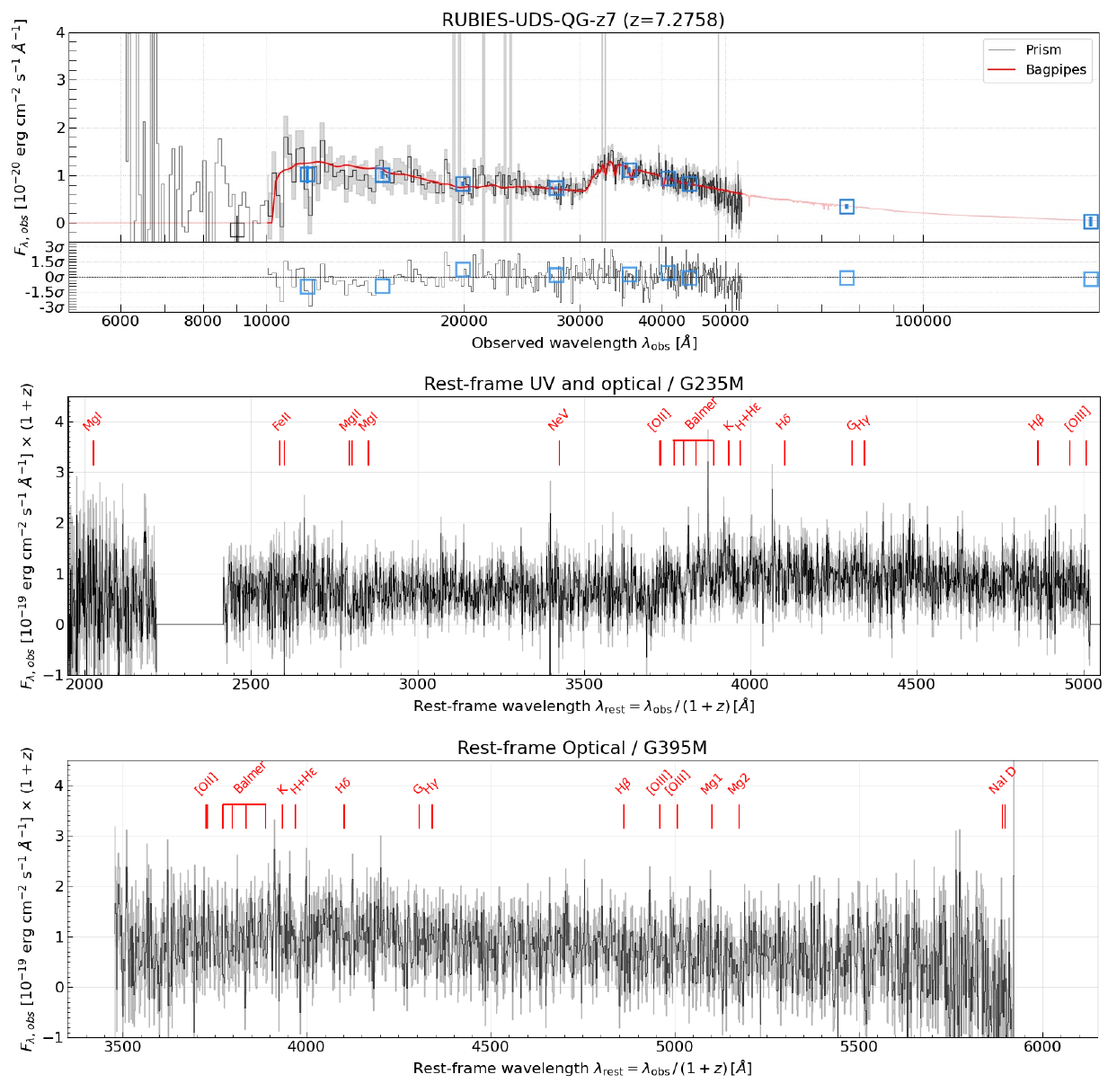}
     \caption{Spectra, photometry, and SED modeling of \rubies\ at $z=7.2758\pm0.0011$. \textit{Top:} Photometry (blue squares), the JWST/NIRSpec PRISM spectrum (black line), and its uncertainty (gray line) that we jointly modeled with \textsc{Bagpipes} (red line) in the observed frame. The normalized residuals of the modeling are shown in the lower inset.
     \textit{Center:} Rest-frame UV spectrum covered by the extended G235M grating. 
     \textit{Bottom:} Rest-frame optical spectrum covered by the G395M grating. We mark in red the expected location of emission and absorption lines at the derived systemic redshift, $z_{\rm spec}=7.2758$, as labeled. The flux densities in the bottom panels have been rescaled by $(1+z)$ to conserve the energy.}
    \label{fig:sed_rubies}
\end{figure*}
\section{Data}

\subsection{Imaging \ }
\label{sec:imaging}
\snhope\footnote{The spectroscopic follow-up was primarily designed to characterize the multiply lensed supernova SN H0pe presented in \cite{frye_2024}. For our target, we adopt the name it has in this work.} lies behind the galaxy cluster PLCK G165.7+67.0 at $z=0.35$ (G165 in brief). This field was imaged as part of the JWST Prime Extragalactic Areas for Reionization and Lensing Science (PEARLS) guaranteed time observations (Program ID, PID, 1176, PI: R. Windhorst). Follow-up Near Infrared Camera (NIRCam; \citealt{rieke_2023}) imaging and Near Infrared Spectrograph (NIRSpec, \citealt{jakobsen_2022, ferruit_2022}) spectroscopy was acquired in a JWST disruptive Director's Discretionary Time program (PID 4446; PI: B. Frye; see \citealt{frye_2024} for the analysis of all JWST observations in this field). This field also benefits from ample ancillary data, such as infrared imaging with the Wide Field Camera 3 (WFC3) onboard the \textit{Hubble} Space Telescope (HST) \citep[GO-14223; PI: B. Frye,][]{frye_2019}, Very Large Array (VLA) observations \citep{pascale_2022}, \textit{XMM-Newton} observations (AO22, \#92030; PI: B. Frye), and large ground-based telescope observations (see \citealt{frye_2024, kamieneski_2024} for a full description; \citealt{canameras_2018_snhope} for previous work on this field). For the purpose of this work, we utilized the available \jwst\ and \hst\ imaging in 8 bands covering the 0.9-4.4 $\mu$m range. The images were retrieved from the DAWN \jwst\ Archive (DJA)\footnote{\url{https://dawn-cph.github.io/dja/index.html}} and homogeneously reduced with \textsc{grizli} \citep{brammer_2023} as described in \cite{valentino_2023}. The photometry was extracted from a combined image (mosaic version v7.0) of the available NIRCam long-wavelength filters in circular apertures (0.5" diameter) with the pythonic version of SourceExtractor \citep{bertin_1996, barbary_2016} and corrected to total within an elliptical Kron aperture \citep{kron_1980}. The aperture correction was computed on the combined long wavelength image and applied to all bands \citep{valentino_2023}. Moreover, we corrected for the minimal extinction of the Milky Way ($E(B-V)=0.0165$) based on the maps in \cite{schlafly_2011}. For reference, the observed magnitude in the F200W filter ($m_{\rm F200W,AB}=23.34\pm0.05$) is consistent with previously reported estimates \citep{frye_2024, wu_2025}. Finally, we checked for the gravitational lensing effect of the cluster at the location of \snhope, finding a mild magnification correction of $\mu=2.0\pm0.2$ in the recent maps by \cite{kamieneski_2024} -- without strongly distorting the shape.\footnote{We note that this target is outside the field of view of the lensed image constraints used in the models \citep{frye_2024}.}

In addition to optical and near-infrared imaging, we obtained coverage of \snhope\ at 1.3 mm (260 $\mu$m rest-frame) with the NOrthern Extended Millimetre Array (NOEMA; ID: W23CU001, PIs: G\'{o}mez-Guijarro \& Valentino). The target was observed for 11.6 hours from May to September 2024 in D configuration. 
The frequency tunings are 212.5--220.0~GHz in the lower sideband and 228--235.5~GHz in the upper sideband.
We reduced and calibrated the data using the GILDAS pipeline at the IRAM headquarters in Grenoble. We produced $uv$ visibility tables and performed analysis in the $uv$ space, following the methods described in \cite{Jin2019alma,Jin2022,Jin2024cosbo7}.
The clean continuum image was produced using the GILDAS HOGBOM clean routine.
The final data product reaches a continuum sensitivity of $33~\mu$Jy at a central frequency of 224~GHz. 
The resulting synthesized beam is $1\farcs85\times1\farcs80$, and thus the source is unresolved.
We do not detect the dust continuum emission from \snhope. We thus placed a $3\sigma$ upper limit on its observed flux density of $\mu F_{\rm 260\,\mu m,\, rest}\leq0.1$ mJy (over a beam, considering that the source is unresolved). We show in Fig. \ref{fig:rgb_noema} a red-green-blue (RGB) image of \snhope\footnote{The image was produced with \textsc{Trilogy} by D. Coe, \url{https://github.com/dancoe/Trilogy}}, where we overlay the contours of the dust continuum emission recorded with NOEMA.

\rubies\ falls in the footprint of the Public
Release IMaging for Extragalactic Research program (PRIMER; PID 1837, PI: J. Dunlop) in the UDS field. In this case, we made use of the available \jwst\ NIRCam and Mid-Infrared Instrument (MIRI) imaging (v7.2) available on DJA and covering the $0.9-18\,\mu$m interval. For consistency with the analysis in \cite{weibel_2025}, we adopted their total photometry, extracted in 0\farcs32 diameter apertures in point-spread-function- (PSF-)matched images, then corrected for aperture effects and the extinction of our Galaxy. The procedure is detailed in \cite{weibel_2024a}, to which a custom extraction of the MIRI photometry in 0\farcs5 diameter apertures in the F770W and F1800W images has been added. We note that following a procedure similar to that employed for \snhope\ returns fluxes $\sim20$\% fainter without introducing any significant color differences, thus not appreciably affecting the results of this paper. 

\subsection{Spectroscopy}
\label{sec:spectroscopy}
\jwst/NIRSpec Micro-Shutter Array (MSA) spectra of sources in the G165 cluster field were obtained with the PRISM/CLEAR ($0.7-5.3\,\mu$m, $R\approx20-300$), medium resolution gratings G140M/F100LP ($0.9-1.8\,\mu$m, $R\approx1000$), and G235M/F170LP ($1.6-3.2\,\mu$m, $R\approx1000$) on April 22, 2023 (PID 4446, PI: B.~Frye). The science exposure times were 4420s, 6696s, and 919s for G140M/F100LP, G235M/F170LP, and the PRISM/CLEAR observations, respectively \citep{frye_2024}. \snhope\ was among the targeted sources. The spectra were acquired with a three-microshutter configuration and a three-nod nodding pattern. Also in this case, we retrieved the spectra from the DJA (version v3). These were reduced with the \textsc{msaexp} pipeline \citep[][]{msaexp} along the same lines detailed in \cite{heintz_2025} and \cite{degraaff_2024_rubies}. The current public version of \textsc{msaexp} includes updated reference files and improves on the absolute and color-dependent flux calibration and the bar shadow correction. The spectra were optimally extracted \citep{horne_1986} and their noise budget conservatively scaled up such that the residuals of a wavelength-dependent polynomial model subtracted to the spectrum are normally distributed, as expected for pure background integrations (although localized noise correlation due to the line spread function could induce deviations from this assumption). This corresponds to a $\sim25$\%, $25$\%, and $45$\%  noise median increase for the PRISM, G140M, and G235M spectra. Finally, we anchored the spectra to the total photometry as derived in Sect. \ref{sec:imaging} to correct for residual slit losses. We applied a fourth-order polynomial correction to the spectra, which resulted in slowly varying rescaling on the order of 5-40\%, the most extreme correction being at the very blue end of the PRISM spectrum. The results are robust against the choice of alternative scaling functions (spline and Chebyshev polynomials) and the degree of the polynomial (a simple constant or parabolic functional form already corrects for most of the offset between the spectra and the total photometry). The photometry and spectra of \snhope\ are shown in Fig. \ref{fig:sed_snhope}.\\

NIRSpec/MSA spectra of \rubies\ were obtained on July 25, 2024, with PRISM/CLEAR and G395M/F290LP as part of the Red Unknowns: Bright Infrared Extragalactic Survey survey (RUBIES, PID 4233, PIs: A. de Graaff \& G. Brammer; \citealt{degraaff_2024_rubies}). Both spectra were acquired with 2880s combined integrations with a three-microshutter configuration and a three-nod nodding pattern. These spectra were reduced following the same steps as detailed above. The photometry and PRISM spectrum of \rubies\ are shown in Fig. \ref{fig:sed_rubies}. The G395M spectrum does not show any appreciable absorption features at high enough S/N \citep{weibel_2025}, which resulted in relatively loose initial redshift constraints ($z=7.29\pm0.01$).

On July 23, 2024, a G235M/F170LP spectrum was also obtained as part of the ``DeepDive'' survey (PID 3567, PI: F. Valentino; Ito et al. in prep.). A similar nodding pattern was used also in this case, but leaving three extra shutters open for an optimal background subtraction. The total integration time was of $10,590$ s.
The G235M/F170LP grating was processed with a customized version of \textsc{msaexp} (v4) that allowed us to recover the whole wavelength range permitted by the combination of the F170LP long-pass filter and projection on the detector (Brammer et al. in prep.). The distributed JWST pipeline automatically cuts the red end of the grating spectra to avoid contamination of the primary first spectral order with higher orders, which appear at predictable locations and intensities \citep{jakobsen_2022}. However, by extending the \texttt{wavelengthrange}\footnote{\url{https://jwst-pipeline.readthedocs.io/en/latest/jwst/references_general/wavelengthrange_reffile.html}} pipeline reference file and calibrating the intensities of the second and third order spectra (which include the full response of the telescope, the shape of the filters, and flat-fielding, all summarized in a revised sensitivity curve), the full spectrum recorded on the detector can be recovered. The sensitivity curve extending beyond the nominal coverage of the filter (1.6--3.2$\,\mu$m) was derived from the wavelength and flux commissioning and calibration programs COM PIDs 1125 (PI: J. Muzerolle Page) and 1128 (PI: N. Luetzgendorf) and CAL PID 1538 (PI: K. Gordon). This allowed us to extend the spectrum up to $4.2\,\mu$m, now crucially covering the full Balmer break (Fig. \ref{fig:sed_rubies}). The normalization and continuum shape of the extended G235M/F170LP spectrum are consistent with those of the G395M/F290LP spectrum in the overlapping regions without introducing any further corrections. Given the intrinsic red spectrum of the target and the decreasing transmission of the G235M disperser at $\lambda \gtrsim2.3\,\mu$m, the uncertainties on the flux calibration after modeling the contamination of the higher order spectra are comparable with the errors on the standard absolute calibration. In addition, the extended G235M spectrum has an approximately $1.7$  times higher spectral resolution than G395M at a given wavelength where they overlap. Finally, to avoid spurious systematics when comparing our analysis with that of \cite{weibel_2025}, we anchored both the extended G235M and the G395M spectra to their PRISM data. In this case, the median noise rescaling amounts to $10$\% and $50$\% for the PRISM and G235M, respectively, and no appreciable difference for G395M\footnote{At this point, the reader might wonder about the extension of the spectra of \snhope. We briefly describe their content and limitations in Appendix \ref{app:extended_snhope}.}. 

\begin{table*}[]
    \centering
    \caption{Measurements and physical properties.}
    \begin{tabular}{lcc}
    \toprule
    \toprule
       Property &  \snhope\ & \rubies\ \\
    \midrule
    R.A. [deg] & 171.8236199 & 34.4296173 \\ 
    Decl. [deg] &  42.4696455 & -5.1122962\\
    $z_{\rm spec}$ & $4.1061\pm0.0003$ & $7.2758\pm0.0011$\\
    $\mu$ & $2.0\pm 0.2$ & $1$ \\
    $\mu$ \neiii$\lambda$3870 [$10^{-19}$ cgs]& $2.4^{+2.8}_{-2.4}$ & $-$\\
    $\mu$ \oii$\lambda$3729 [$10^{-19}$ cgs]& $<6.6$ & $-$\\
    $\mu$ \hb$\lambda$4863 [$10^{-19}$ cgs]& $8.0^{+3.7}_{-3.8}$ & $-$\\
    $\mu$ \oiii$\lambda$5008 [$10^{-19}$ cgs]& $8.9^{+3.9}_{-4.2}$ & $-$\\
    $\mu$ $m_{\rm F200W,AB}$ [mag]& $23.34\pm0.05$& $26.31\pm0.06$\\ 
    $\mathrm{log}(M_\star/M_\odot)$ & $10.57^{+0.02}_{-0.02}$ & $10.13^{+0.07}_{-0.02}$\\
    $A_{\rm V}$ [mag]&                    $0.40^{+0.04}_{-0.04}$ & $0.50^{+0.06}_{-0.10}$\\
    $\rm SFR_{100\rm Myr}$ [\myr]&     $10.7^{+0.1}_{-0.1}$ & $33^{+93}_{-32}$\\
    $\rm SFR_{10\rm Myr}$ [\myr]&     $9.0^{+0.1}_{-0.1}$ & $<2$\\
    $\rm SFR_{\rm IR,\,8-1000\mu m}$ [\myr]&     $<12$ & $-$\\
    $\rm SFR_{[OII]}$ [\myr]&   $<0.4$ & $-$\\  
    $\rm SFR_{H\beta}$ [\myr]&  $1.5^{+0.7}_{-0.7}$ & $-$ \\  
    $\rm log(sSFR_{\rm 100Myr}$/yr$^{-1}$)&  $ -9.54^{+0.05}_{-0.05}$ & $-8.7^{+0.6}_{-1.9}$ \\  
    \sigmastar\ [\kms]&             $255^{+39}_{-42}$ & $-$ \\
    $\Delta v_{\rm off}$ [\kms]&    $-182 \pm 50$ & $-169\pm46$\\  
    $R_{\rm maj}$ [pc]&            $262^{+17}_{-15}$ & $209^{+33}_{-24}$\\  
    $n_{\rm Sersic}$ &              $6.5^{+0.6}_{-0.5}$ & $2.4^{+1.5}_{-0.9}$\\  
    $\mathrm{log}(M_{\rm dyn}/M_\odot)$ &   $10.18^{+0.13}_{-0.16}$ & $-$\\  
    \midrule
    $\mathrm{log}(N({\rm FeII})/\mathrm{cm}^{-2})$&   $15.00 \pm 0.19$ & $-$\\
    $\mathrm{log}(N({\rm MgII})/\mathrm{cm}^{-2})$&   $15.35 \pm 0.74$ & $\sim17.9$\\
    $\mathrm{log}(N({\rm NaI})/\mathrm{cm}^{-2})$&   $12.15 \pm 0.59$ & $-$\\
    $\mathrm{log}(N(\rm H_{\rm Fe})/\mathrm{cm}^{-2})$&   $21.19 \pm 0.19$& $-$\\
    $\mathrm{log}(N(\rm H_{\rm Mg})/\mathrm{cm}^{-2})$&   $20.57 \pm 0.74$& $\sim23.1$\\
    $\mathrm{log}(N(\rm H_{\rm Na})/\mathrm{cm}^{-2})$&   $19.79 \pm 0.59$ & $-$\\
    $\dot{M}_{\rm out}$ [\myr] (Fe, Mg, Na)& $(5^{+5}_{-3},\,1^{+7}_{-1},\,0.2^{+0.8}_{-0.2})$ & ($-,\sim269,-)$\\ 
    $\eta$ (Fe, Mg, Na)&            $(0.2^{+0.3}_{-0.1},\,0.1^{+0.4}_{-0.1},\,0.01^{+0.04}_{-0.01})$ & ($-,\sim12,-)$\\ 
    \bottomrule
  \end{tabular}
  \tablefoot{Upper limits at $3\sigma$. The \neiii, \oii, \hb, and \oiii\ fluxes refer to emission lines obtained after subtracting the underlying stellar continuum. \mdotout\ and $\eta$ are computed assuming $v_{\rm out}=\Delta v_{\rm off}$ and $\rm SFR_{100Myr}$ (Sect. \ref{sec:outflow_properties} for a discussion of alternative options).}
    \label{tab:results}
\end{table*}

\section{Methods}
\subsection{Spectrophotometric modeling}
\label{sec:modeling}

\subsubsection{\snhope}
Initially, we re-derived the systemic redshift of \snhope\ by modeling the stellar absorption features in the rest-frame optical probed by the G235M grating (Fig. \ref{fig:sed_snhope}). We did so with the Penalized PiXel Fitting (pPXF) code \citep[][]{cappellari_2004} and following an approach similar to that in \cite{cappellari_2023}. We used models from the Flexible Stellar Population Synthesis (FSPS) library \citep[][]{conroy_2010} normalized around the rest-frame optical V-band. We ran the code twice: we initially masked regions potentially contaminated by emission lines, then identified and excluded $3\sigma$ outlier pixels, and reran the modeling. We allowed only for multiplicative fourth-order polynomials as correction factors to the overall shape of the spectrum, but this choice did not affect the robustness of the redshift solution ($z_{\rm spec} = 4.1061\pm0.0003$,  which is slightly lower than, but consistent with, the $z=4.1076\pm0.0023$ first reported in Table~2 of \citealt{frye_2024}) and of the stellar velocity dispersion estimate. We derived a value of $\sigma_\star = 255^{+39}_{-42}$ \kms\ by modeling the stellar absorption features in the spectrum, excluding deep Balmer lines (H$\beta$, H$\gamma$, and H$\delta$) and other portions of the wavelength range possibly impacted by strong emission lines (\oiii\ and \oii). We corrected this estimate for the wavelength-dependent instrumental resolution, which we assumed was a factor of $1.3$ higher than the nominal resolving power \citep{degraaff_2024_rubies}. Finally, we attempted to add an ionized gas component typically responsible for the emission of bright rest-frame optical emission lines (e.g., Balmer lines, \oii, and \oiii), but we did not retrieve any meaningful detections. We thus place upper limits on the emission lines fluxes covered by the G235M grating by assuming a line width identical to that of the stellar absorption\footnote{See Appendix \ref{app:extended_snhope} for an attempt to measure emission lines in the extended G235M spectrum.} (see \citealt{ubler_2024} and references therein for a recent discussion on this assumption). These values are reported in Table \ref{tab:results}. The uncertainties are derived by bootstrapping the spectrum 500 times and repeating the modeling procedure.\\

Once the systemic redshift was determined, we simultaneously modeled NIRCam and WFC3 photometry and the combined NIRSpec grating data using \textsc{Bagpipes} \citep{carnall_2018,carnall_2020}. We masked spectral regions potentially contaminated by the absorption of the interstellar medium (iron and magnesium lines in the rest-frame UV, the calcium and sodium doublets in the optical). We adopted \cite{bruzual_2003} stellar population synthesis models, the parameterization of the dust attenuation law in \citet[with fixed UV bump amplitude, $B=1$, and slope deviation from a \citealt{calzetti_2000} law, $\delta=-0.2$, for a shape similar to that in \citealt{kriek_2013}]{salim_2018}, and a default \cite{kroupa_2002} initial mass function (IMF). We adopted a double-power law star formation history (SFH), which offers enough flexibility to capture a fast rise and decline of the star formation rate (SFR) in the past. We fixed the value of the stellar velocity dispersion to $\sigma_\star=255$ \kms.
The adopted priors and their shapes are collected in Table \ref{tab:priors} and shown in Fig. \ref{app:fig:priors}.
\begin{figure}
    \centering
    \includegraphics[width=\columnwidth]{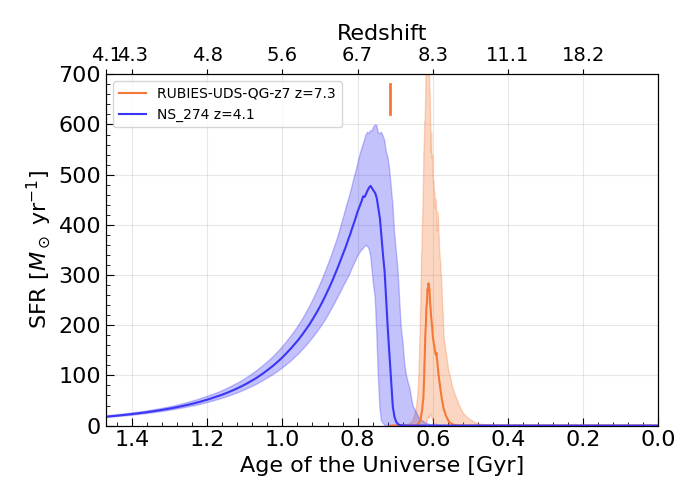}
    \caption{SFHs of our targets. We show the best-fit double power-law SFHs of \rubies\ (orange) and \snhope\ (blue) and their uncertainties. The age of the Universe is truncated at the observed redshift of \snhope. The orange segment marks the observed redshift of \rubies.}
    \label{fig:sfh}
\end{figure} 
The fitted data and best model are shown in Fig. \ref{fig:sed_snhope}, while we report best-fit parameters of interest in Table \ref{tab:results}. A corner plot with the posterior distributions of such parameters is shown in Appendix \ref{app:fig:corner}. We tested these results against the use of looser priors, different parameterizations of the SFHs (e.g., lognormal and delayed $\tau$-models), and a different code \citep[Fast++;][]{schreiber_2018b}. We also attempted to model the photometry, grating, and PRISM independently or combined. Finally, we explored the effect of the imposed priors on the final estimates of interesting physical quantities at test. We found the best-fit estimates to be consistent across this range of tests and robust against the choice of the priors. Predictably, the range of parameter estimates obtained under different assumptions is larger than what implied by the statistical uncertainties with a single set of priors and premises (see \citealt{pacifici_2023} for a comparative study of systematic effects in spectral energy distribution modeling). As an example, these tests provide \mstar\ estimates varying within a $\sim0.2$ dex range encompassing our fiducial values in Table \ref{tab:results}.\\

Accounting for the magnification effect ($\mu=2.0\pm0.2$), we estimate an intrinsic stellar mass of $\sim4\times10^{10}$\msun. The spectral energy distribution (SED) modeling suggests that \snhope\ underwent a short and intense (peak $\mu$SFR $\sim500$ \myr) growth spurt approximately $\sim500$ Myr prior to the time of observations, leaving behind some modest level of SFR over the last 100 Myr (intrinsic $\rm SFR_{\rm 100\,Myr}=10.7^{+0.1}_{-0.1}$ \myr) and 10 Myr ($\rm SFR_{\rm 10\,Myr}=9.0^{+0.1}_{-0.1}$ \myr; Fig. \ref{fig:sfh}).
We also constrain the dust obscured SFR by rescaling a modified black body with dust temperature $T_{\rm dust}=35$ K and $\beta$ slope of $1.8$ \citep{witstok_2023} to the upper limit on the continuum emission at 260 $\mu$m rest-frame, while accounting for the effect of the cosmic microwave background \citep{dacunha_2013}. The resulting $\rm SFR_{IR,\,8-1000\mu m}$, using the conversion in \cite{kennicutt_1998} adapted for our assumed IMF and corrected for the magnification factor, is $<12$ \myr\ (at the $3\sigma$ level). This value is consistent with the estimate from the modeling of the shorter wavelength SED, also characterized by a low estimate of $A_{\rm V}$. 
The absence of strong \oii, \oiii, and \hb\ emission further constrains the more recent SFR on shorter timescales of $10$ Myr. It also implies the absence of a a bright AGN radiating efficiently (e.g., \oiii/\hb$=1.1^{+1.2}_{-0.5}$; but see Appendix \ref{app:extended_snhope} and \citealt{wu_2025} for caveats and limits on faint and radiatively inefficient AGN activity). Table \ref{tab:results} reports the constraints on SFR from the emission lines assuming the conversion in \cite{kennicutt_1998} for \hb\ and \cite{kewley_2004} for \oii, correcting for the dust attenuation for the stellar continuum from the SED modeling and for the assumed IMF.
These constraints place \snhope\ in a post-starburst phase, as also supported by the deep stellar Balmer absorption lines. Estimates of the Lick indices ($\rm H\gamma_{\rm A}=6.91\pm0.80$ \AA\ and $\rm H\delta_{\rm A}=7.83\pm0.75$ \AA, \citealt{worthey-ottaviani_1997}) and of the $D_n4000$ index ($D_n4000=1.25\pm0.02$, \citealt{balogh_1999}), computed following the procedure detailed in \citealt{gallazzi_2014}, are indeed typical of post-starburst systems in the local Universe. 
\begin{figure}
    \centering  \includegraphics[width=\columnwidth]{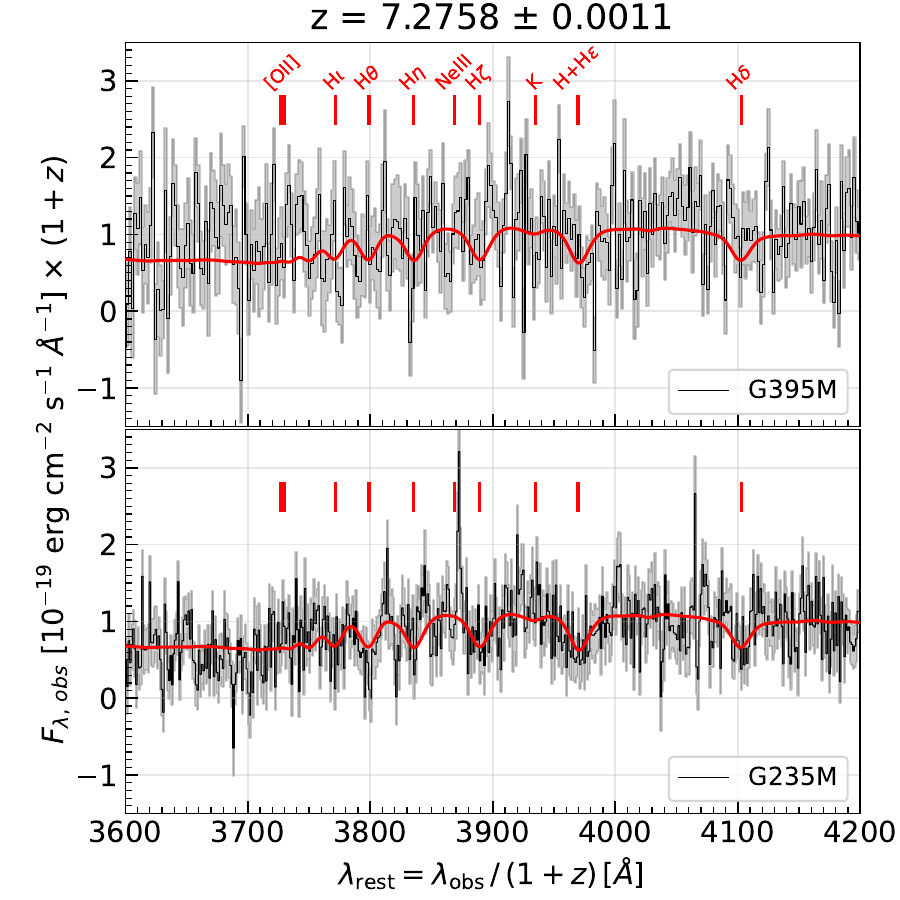}
    \caption{Zoomed-in view of the Balmer break region of \rubies\ in the G395M spectrum from \citet[][\textit{top}]{weibel_2025} and our extended G235M spectrum (\textit{bottom}). The black and gray lines show the data and their uncertainties. The red lines indicate the best-fit model (Sect. \ref{subsec:rubies_zspec}). The red ticks indicate the most notable absorption features, as labeled.}
    \label{fig:rubies_zspec}
\end{figure} 

\subsubsection{\rubies}
\label{subsec:rubies_zspec}
The extended G235M spectrum shows absorption features that are not visible in the G395M spectrum (Fig. \ref{fig:rubies_zspec}). While the depth of the spectrum does not allow for a robust estimate of the stellar velocity dispersion from these lines, a collective model of their wavelengths tightly constrains the redshift. To combine the information from both the G235M and G395M spectra at their respective resolutions, we proceeded as follows. We ran pPXF with a similar setting as for \snhope\ on the extended G235M spectrum. We then took the  best-fit template at its intrinsic resolution and simultaneously fitted it to both the G235M and G395M spectra, convolving the model to their corresponding resolution curves. We allowed for a normalization rescaling for each spectrum, but the shape of the template was fixed. The remaining free parameters were the redshift and the velocity dispersion. The fitting procedure was implemented using \textsc{emcee} \citep{foreman-mackey_2013}, where we assumed a Gaussian likelihood and flat priors. The best-fit model is shown in Fig. \ref{fig:rubies_zspec}, corresponding to $z=7.2758\pm0.0011$, which is consistent with the PRISM-based determination. The corner plot with the posterior distributions of the free parameters is shown in Fig. \ref{app:fig:zspec_corner}. As noted, the velocity dispersion is poorly constrained, which is why it was not considered any further in the analysis.\\

For consistency with the work by \cite{weibel_2025}, we opted for modeling the PRISM spectrum and photometry with \textsc{Bagpipes} using the same set of assumptions as for our fiducial model of \snhope, but fixing the redshift to our revised estimate. The best-fit parameters (Table \ref{tab:results}) and the overall shape of the SFHs are broadly consistent with those reported in \cite{weibel_2025}. The modeled data and the best-fit models are shown in Fig. \ref{fig:sed_rubies}. 
Also in this case, the modeling suggests that \rubies\ experienced a short and vigorous main formation burst (peak SFR $\sim200$ \myr) approximately $\sim150$ Myr prior to the observed redshift, immediately followed by an abrupt suppression of the SFR ($\rm SFR_{100\,Myr}=33^{+93}_{-32}$ \myr and $\rm SFR_{10\,Myr}<2$ \myr at $3\sigma$; Fig. \ref{fig:sfh}). As noted in the past \citep{deGraaff_2025}, the unconstrained stellar metallicity does have an impact on the reconstructed SFH and SFR estimates. In this case, the preferred solution ($Z\sim0.5Z_\odot$) is in between the low- ($\sim0.1Z_\odot$) and high-$Z$ ($\sim Z_\odot$) models in \cite{weibel_2025}. On shorter timescales, the upper limit on the emission lines set a constraint on the SFR of $\lesssim6$\myr \citep{weibel_2025}. 

\begin{figure*}
    \centering     
    \includegraphics[width=\textwidth]{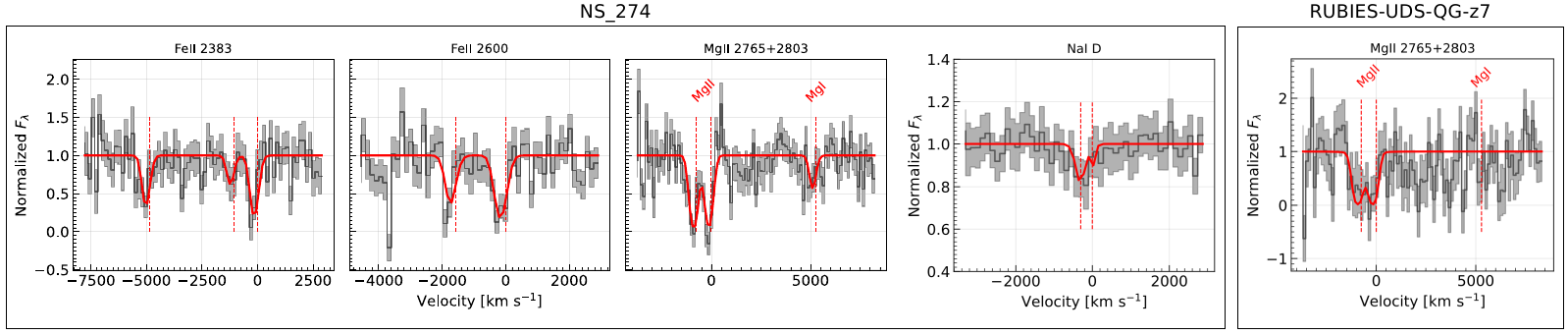}     
    \caption{Blueshifted absorption features. Left panels: Continuum-normalized spectrum (black line) and its uncertainties (gray line) around FeII, MgII+MgI, and NaI D lines for \snhope. Rightmost panel: G235M spectrum around MgII+MgI for \rubies. The best-fit model with \textsc{VoigtFit} is marked by the solid red line. The systemic velocity for each line is marked by dashed red lines.}
    \label{fig:restuv}
\end{figure*}

\subsection{Modeling the rest-frame UV and optical absorption features tracing outflowing neutral gas}

A few rest-frame UV absorption lines tracing the presence of neutral gas from low-ionization elements in galaxies (\mgii\ doublet at $\lambda\lambda2796,\,2803$; \feii\ at $\lambda\lambda2586,\,2600$; $2374,\,2383$) are robustly detected in the G140M spectrum of \snhope\ (Fig. \ref{fig:restuv}). Absorption features at similar redshifts are also at least tentatively detected (e.g., \mgi\ at $\lambda$2853). In the rest-frame optical covered by the G235M grating, we also robustly detect the Na\,{\sc i} $\lambda\lambda$5891, 5897 doublet in absorption (NaI D, although blended at this resolution), indicating an additional dense component in the neutral gas-phase.
Common to all these features is a blueshift compared to the systemic redshift derived from the well-detected stellar absorption features, indicating a substantial column of outflowing gas in the line of sight. In the case of \rubies, given the limited S/N and coverage, we are only able to robustly detect the presence of \mgii\ (Fig. \ref{fig:restuv}).\\

To explore the properties of the outflowing gas, we consistently modeled these absorption features with \textsc{VoigtFit} \citep{krogager_2018_voigtfit}\footnote{\url{https://github.com/jkrogager/VoigtFit}} following the approach in \cite{heintz_2023_dtg}. The code models the lines with Voigt profiles at the delivered spectral resolution and returns the column densities, $N,$ and broadening parameters, $b,$ due to thermal and turbulent motions for an arbitrary number of velocity components. Since absorption from stars and the interstellar medium (or even emission from the latter) can arise at the systemic redshift \citep{sugahara_2017, man_2021, davies_2024}, when modeling the lines in the spectra of \snhope, we implemented a two-component model, with one at fixed $z=z_{\rm spec}$. We also ran the code iteratively: a first pass with free broadening parameter and velocity offsets, $\Delta v_{\rm off}$, to determine the latter, followed by a second iteration at fixed parameters to derive the column densities of the outflowing component. However, given the available resolving power ($R\approx1000$) and depth, the final  estimates of $N$ are affected by significant uncertainties (Table \ref{tab:results}). 
The same conditions also limit more refined modeling of stellar population in the rest-frame UV \citep[][]{maltby_2019}, which we thus leave for future work with better data quality (but see \citealt{wu_2025} for an attempt). When simultaneously modeling the Fe, Mg, and Na transitions, we tied their broadening parameter and velocity structure, assuming that they physically trace the bulk of the neutral gas-phase in galaxies. Considering that the ionization potentials of the elements considered here (Na 5.1 eV, Fe 7.9 eV, Mg 7.6 eV) are comparable and lower than that of hydrogen (13.6 eV), this assumption is physically motivated. 
Finally, we note that, in principle, $\mathrm{log}(N(\mathrm{MgII}))$ might be underestimated due to saturation of the line core. Given the limited information in the spectra of \rubies, we did not attempt a two-component model and fixed $b=100$ \kms, which is consistent with the results for \snhope. We caution the reader that this might lead to an underestimate of the velocity offset and of $\mathrm{log}(N)$. Deeper and higher resolution spectra will help address several of these issues in the future.\\

The best-fit models obtained with \textsc{VoigtFit} are shown in Fig. \ref{fig:restuv}. We derive an average velocity offset of the absorbing gas of $\Delta v_{\rm off} = -182 \pm 50$ \kms\ and $-169\pm46$ \kms\ for \snhope\ and \rubies, respectively, with tails extending up to a few hundred \kms. In the case of \snhope, the velocity structure is derived from the joint modeling of the \mgii, \mgi, \feii, and NaI lines mentioned above, and the offset is consistent with the estimate first reported in \cite{wu_2025}. However, the wavelength shifts of each line are consistent. These results indicate that the dominant neutral gas column is associated with the outflowing gas. The estimated column densities are reported in Table \ref{tab:results}. We converted $\mathrm{log}(N({\rm FeII}))$, $\mathrm{log}(N({\rm MgII}))$, and $\mathrm{log}(N({\rm NaI}))$ into $\mathrm{log}(N_{\rm H})$, assuming a gas-phase metallicity and empirical depletion patterns, under the same assumptions detailed in \cite{wu_2025}:
\begin{equation}
    \mathrm{log}(N_{\rm H}) = \mathrm{log}(N_{\rm X}) - \delta_{\rm depl,\,X} - \mathrm{log}(N_{\rm X}/N_{\rm H})_\odot - \mathrm{log}(Z/Z_\odot),
\end{equation}
where $X=\rm Fe,\,Mg, and\,Na$; $\delta_{\rm depl,\,X}$ is the metal depletion onto dust ($\delta_{\rm Fe}=-1.7$, $\delta_{\rm Mg}=-0.8$ are the median values in the Galactic disk \citep{jenkins_2009}; and $\delta_{\rm Na}=-0.95$ is the canonical estimate for the Milky Way from \citealt{savage-sembach_1996}); $\mathrm{log}(N_{\rm X}/N_{\rm H})_\odot$ are the solar abundance patterns for Fe (-4.49), Mg (-4.42), and Na (-5.69, \citealt{savage-sembach_1996}); and $Z$ the metallicity of the targets. We included a neutral fraction correction of a factor of ten for Na\,{\sc i}, but this number could be substantially higher in the extreme conditions of a recent burst of star formation or AGN activity at high redshifts  \citep{veilleux_2020}. We do not apply any ionization corrections for Fe and Mg, assuming that the singly ionized ions are the dominant population. This allows us to draw a direct comparison with the analysis in \cite{wu_2025} without introducing major systematics. For reference, we assume $Z=Z_\odot$. The values from our spectrophotometric modeling are consistent with it, but they are a source of major uncertainty. This is particularly true for \rubies, whose high redshift would suggest a metallicity lower than and an abundance pattern different from solar \citep{weibel_2025}.\\

Finally, we note that, at this stage, the velocity offset can in principle be interpreted as due to a genuine neutral gaseous outflow or from a cloud intervening gas along the line of sight. However, this second possibility is disfavored by the absence of credible galaxies associated with the putative absorbers along the line of sight in the surrounding of our targets. Moreover, the implied high column densities (Sect. \ref{sec:outflow_properties}) require a non-negligible level of metal enrichment, and thus underlying stellar mass, which is not detected (although low surface brightness galaxies that fall below the flux limit cannot be ruled out).

\subsection{Structural properties}
\label{sec:structure}

Morphological properties for \rubies\ were derived in \cite{weibel_2025} using the Bayesian code \textsc{pysersic} \citep{pysersic}\footnote{\url{https://github.com/pysersic/pysersic}}. The authors retrieve an effective semimajor axis of $R_{\rm maj} = 209^{+33}_{-24}$ pc for a rather round shape (axis ratio of $R_{\rm min}/R_{\rm maj}=0.89^{+0.08}_{-0.14}$, where $R_{\rm min}$ is the semiminor axis) and a relatively low, but poorly constrained, \cite{sersic_1968} index of $n=2.4^{+1.5}_{-0.9}$. We followed a similar procedure and modeled the emission of \snhope\ in the F200W band using the same code. The choice of the band is a compromise between mapping relatively long wavelengths ($\sim4000$\AA\ rest frame) at the highest possible resolution (sampled with a $0\farcs02$ pixel scale) and S/N. We built a PSF model by broadening the profile derived with \textsc{WebbPSF} \citep{perrin_2014} to match the width of observed unsaturated stars in the field, as described in \cite{ito_2024}.
We modeled a single \cite{sersic_1968} profile. The source is resolved: we estimate an effective semimajor axis of $R_{\rm maj} = 262^{+17}_{-15}$ pc
and a high $n=6.5^{+0.6}_{-0.5}$ index. We applied an average $\mu^{-0.5}$ correction to the effective size to account for the mild lensing effect. Nevertheless, this correction does not substantially alter the conclusions of this work. The source is slightly elongated (ellipticity $\epsilon = 1 - R_{\rm min}/R_{\rm maj} = 0.49^{+0.02}_{-0.02}$). 
We show the best-fit model and the residuals in Appendix \ref{app:bestfit_modeling}.
These results are consistent with the analysis with \textsc{Galfit} \citep{peng_2002} in \cite{wu_2025}, once the lensing correction is accounted for.\\ 

Combined with the stellar velocity dispersion, we estimated the dynamical mass expected for \snhope\ as\begin{equation}
    M_{\rm dyn} = K(n)\,K(q)\,\sigma_\star^2\, R_{\rm maj} / G,
\end{equation}
where $K(n) = 8.87-0.831n+0.0241n^2$ depends on the S\'{e}rsic index $n$ \citep{cappellari_2006},  $K(q) = (0.87+0.38e^{-3.78(1-q)})^2$ depends on the projected axis ratio $q$ \citep{van_der_wel_2022}, $R_{\rm maj}$ is the semimajor axis, and $G$ is the gravitational constant. After accounting for the lensing effect, $\mathrm{log}(M_\star/M_\odot)=10.57^{+0.02}_{-0.02}$ is higher than, but in overall agreement with, 
$\mathrm{log}(M_{\rm dyn}/M_\odot)=10.18^{+0.13}_{-0.16}$, especially considering the numerous assumptions on the stellar population and dynamical modeling behind each of these calculations and the typical systematic uncertainties ($\sim0.2$ dex on the stellar mass). 

\section{Outflow properties}
\label{sec:outflow_properties}

At this point we had ascertained the presence of \mgii\ (and in the case of NS\_274 also \mgi, \feii, and NaI D) blueshifted absorption features in the spectra of two post-starburst, recently quenched galaxies at high redshifts.\\

How much gas is outflowing from these galaxies and at what rate? Adopting the simple spherical thin-shell model in \cite{wu_2025} and \cite{davies_2024}, we derived the outflowing mass \mout\ and its rate \mdotout\ as
\begin{equation}
\begin{split}
M_{\rm out} & = 1.4\,m_{\rm p}\,\Omega\,N(\mathrm{H})\,R_{\rm out}^2,\\
\dot{M}_{\rm out} & = 1.4\,m_{\rm p}\,\Omega\,N(\mathrm{H})\,R_{\rm out}\,v_{\rm out},
\end{split}
\end{equation}
where $m_{\rm p}$ is the proton mass, $\Omega$ the solid angle subtended by the outflow, $R_{\rm out}$ the shell radius, and $v_{\rm out}$ its velocity \citep{rupke_2005}. We adopt the same approach as in \cite{wu_2025} to directly compare with their results and avoid systematics. We thus assume $\Omega = 0.45\times4\pi$ (based on \citealt{davies_2024}), $R_{\rm out}=2R_{\rm maj}$, and $v_{\rm out}=\Delta v_{\rm off}$ is the average outflow velocity from our fit (Table \ref{tab:results}). For \snhope, we obtain $\mathrm{log}(M_{\rm out}/M_\odot) = 7.4^{+0.2}_{-0.3}$, $6.8^{+0.7}_{-0.9}$, and $6.0^{+0.6}_{-0.7}$ from Fe, Mg, and Na, respectively. The estimate based on Fe is in good agreement with that reported in \citet[$\mathrm{log}(M_{\rm out}/M_\odot) \sim 7.8$]{wu_2025}, while that derived from Mg is $\sim10\times$ larger. The outflow mass from Fe and Mg are in agreement, while that from Na is lower by $\sim1$~dex at face value. This might be due to the fact that the observed transitions, while broadly tracing the neutral hydrogen phase of the ISM, are sensitive to gas conditions, with NaI D preferentially being a proxy of the colder and denser gas.
For \rubies, we derive $\mathrm{log}(M_{\rm out}/M_\odot)\sim9.1$ from Mg, based on a column density of $\mathrm{log}(N(\mathrm{MgII})/\mathrm{cm^{-2}})\sim17.9$. These estimates are more uncertain than those derived for \snhope\ (largely due to the assumption on $b$, the absence of S/N to constrain Fe lines, and a systemic component) and their face values should be taken with a grain of salt. The derived HI column density is remarkably high (Table \ref{tab:results}) and, if confirmed, its effects could be tested in deeper PRISM data around the Lyman limit \citep[e.g.,][]{heintz_2025}.\\

We estimate mass outflow rates of $\dot{M}_{\rm out}\sim0.2-5$ \myr\ and $\sim269$ \myr\ for \snhope\ and \rubies, respectively. The range in \mdotout\ for \snhope\ reflects the difference in $\mathrm{log}(N(\mathrm{H}))$ from Na and Fe. The high column density from Mg dominates this high, albeit uncertain, estimate of \mdotout\ for \rubies. The corresponding mass loading factors are $\eta=\dot{M}_{\rm out} / \mathrm{SFR}\sim0.01-0.2$ and $\sim12$ for \snhope\ and \rubies, where we adopted the SFR averaged over the last 100 Myr for the calculation. The estimates of \mdotout\ and $\eta$ linearly depend on $v_{\rm out}$, and the mass loading factor is also sensitive to the value of the SFR. On the one hand, adopting the alternative assumption of $v_{\rm out} = \Delta v_{\rm off}+2\sigma$, where $\sigma$ is the velocity dispersion of the outflow absorption feature \citep[e.g.,][]{davies_2024}, we would obtain \mdotout\ and $\eta$ estimates $1.8\times$ higher for both targets, given our modeling of the outflow velocity structure with fixed $b=100$ \kms\ and assuming pure Doppler broadening ($b=\sqrt{2}\sigma$). We stress that higher resolution observations are necessary to properly constrain $b$ and break its degeneracy with the column densities. On the other hand, assuming an estimate of SFR averaged on shorter timescales, would have a lesser effect on the estimate of $\eta$ for \snhope\ at fixed $v_{\rm out}$ ($\rm SFR_{100Myr}$ is $\sim20$\% higher than $\rm SFR_{10Myr}$, Table \ref{tab:results}). Instead, it would boost even further $\eta$ for \rubies\ (the $3\sigma$ upper limit of $\rm SFR_{10Myr}<2$ \myr\ would correspond to a lower limit on $\eta$ more than ten times higher than that derived assuming $\rm SFR_{100Myr}$). The choice of a shorter timescale is also consistent with the expected time that an outflow would take to reach our reference scale of $R_{\rm out}=2R_{\rm maj}$ with a constant $v_{\rm out}=\Delta v_{\rm off}$ ($\sim2.5$ Myr for both targets).

\begin{figure*}
    \centering
    \includegraphics[width=\textwidth]{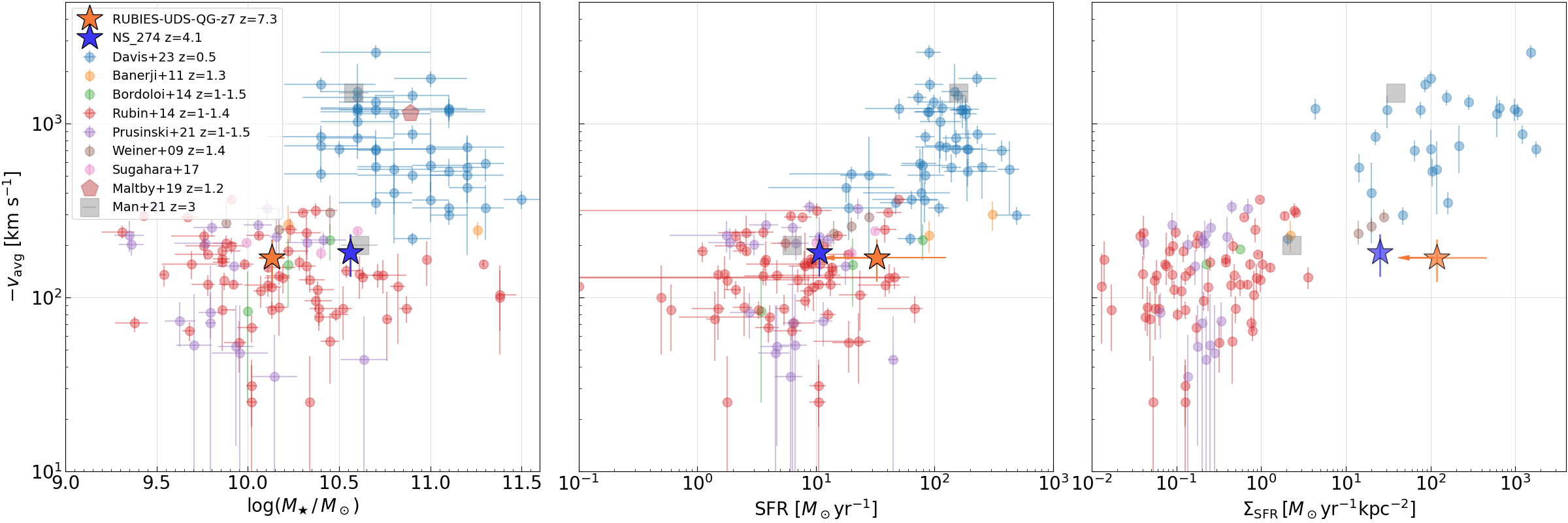}
    \caption{Average outflow velocity as a function of the stellar mass, $\mathrm{SFR}_{\rm 100Myr}$, and surface density of $\mathrm{SFR}_{\rm 100Myr}$ (corrected by the magnification factor for \snhope). Filled stars indicate the targets analyzed in this work. The colored filled circles represent the homogenized literature compilation of \mgii\ outflows across redshifts from \citet[][and references therein]{davis_2023}. The results from the stacking of post-starburst galaxies at $z\sim1.2$ from \cite{maltby_2019} are shown with a red pentagon.
    The locations of two $z\sim3$ recently quenched sources in \cite{man_2021} are marked by gray squares.}
    \label{fig:literature}
\end{figure*}

\section{Discussion}
\label{sec:discussion}

\subsection{Outflow velocities in recently quenched galaxies across time}
In Fig. \ref{fig:literature} we show the average outflow velocities $v_{\rm out}$ as a function of \mstar, SFR, and SFR surface density ($\Sigma_{\rm SFR}=\mathrm{SFR}/(2\pi R_{\rm eff}^2)$) for samples of star-forming and post-starburst galaxies up to $z\sim1.5-2$ based on the large literature compilation in \cite{davis_2023}. 
In this case, we considered outflows predominantly traced by Mg and Fe absorption to minimize the systematics in the comparison (see \citealt{sun_2024} for an analysis of outflows traced by NaI D in local post-starburst galaxies).
Interestingly, in terms of average outflow velocities, the neutral gas outflows in \snhope\ and \rubies\ are broadly consistent with those observed in local star-forming and post-starburst galaxies at fixed \mstar, SFR, or $\Sigma_{\rm SFR}$. The velocity offsets measured in two lensed recently quenched systems at $z\sim3$ in \cite{man_2021} and in the stacked spectra of post-starburst galaxies at $1<z<1.4$ in \cite{maltby_2019} are also shown for reference. In both cases, they are also consistent with the correlations observed at lower redshifts.\\ 

The fact that the average velocities detected in \snhope\ and \rubies\ at high redshift are overall consistent with those observed in nearby post-starburst galaxies seems to suggest that the mechanisms launching them during or shortly after quenching are similar to those operating in the local Universe. This is best captured in the tightest among the correlations, the one between $v_{\rm out}$ and $\Sigma_{\rm SFR}$ \citep{heckman_2015}. In this case, the typically higher SFRs at fixed stellar masses at high redshift are balanced by an increasing compactness, so that the launch of winds via copious star formation occurring in small volumes, as one could expect to detect in early galaxies immediately after quenching, is consistent with that observed in rarer, similarly massive systems at low redshift. 
However, only larger samples of recently quenched galaxies at $z>2-3$ will tell us if this is the case, or if a bona fide evolution of the $v_{\rm out}$ - $\Sigma_{\rm SFR}$ (and \mstar, SFR) relations exists \citep{sugahara_2017}.

We also note that the scaling relations considered here do not include any explicit terms to account for possible AGN activity. Interestingly, samples specifically selected or consistent with post-starburst galaxies shown in Fig. \ref{fig:literature} do not show evident signs of ongoing powerful AGN activity (\citealt{maltby_2019, davis_2023}; the sources in \citealt{man_2021} may host faint Type II AGNs based on their \oiii/\oii\ ratios). According to our analysis, this is the case also for the galaxies analyzed here (but see Appendix \ref{app:extended_snhope} and \citealt{wu_2025}). Taken together, this suggests that ongoing AGN activity is not necessarily required to explain the average outflow velocities of \snhope, \rubies, but also lower-redshift post-starburst galaxies in Fig. \ref{fig:literature} at fixed \mstar, SFR, or $\Sigma_{\rm SFR}$.\\   

\subsection{Outflow masses, loading factors, and energetics}

If the outflows in \snhope\ and \rubies\ show velocities similar to those observed in more local post-starburst systems at fixed \mstar, SFR and $\Sigma_{\rm SFR}$ without invoking any strong AGN contribution, the simple calculations in Sect. \ref{sec:outflow_properties} depict a partially different scenario, at least for \rubies. 

\subsection{Outflows as relics of the star formation process winding down in \snhope}
For \snhope, we find that the mass loading factor $\eta=\dot{M}_{\rm out}/\mathrm{SFR}=0.01-0.2$ and the energy associated with the outflow ($\dot{E}_{\rm out}=0.5\dot{M}_{\rm out} v_{\rm out}^2 \sim 0.2-5\times10^{40}\,\mathrm{erg}\,\mathrm{s}^{-1}$) are consistent with basic expectations for SNe-powered outflows, where the minimum and maximum values of these ranges are derived from the estimates of $N\rm(NaI)$ and $N(\rm FeII)$, respectively (Table \ref{tab:results}). To show this, we adopt the approach in \cite{veilleux_2005}. Based on the Starburst99 \citep{leitherer_1999} models with solar metallicity and a \cite{salpeter_1955} IMF, we expect an injection of energy due to SNe on the order of $\dot{E}_{\rm SN}=7\times10^{41}\times \mathrm{SFR_{\rm 100Myr}}=(7.5\pm0.1)\times10^{42}\,\mathrm{erg}\,\mathrm{s}^{-1}$ and a mass outflow rate $\dot{M}_{\rm SN}=0.26 \times \mathrm{SFR_{\rm 100Myr}}=(2.8\pm0.1)$ \myr. Even neglecting a simple linear scaling to match the IMF of our choice and assuming coupling with the ISM on the order of $\sim10$\% or less, the injected energy is broadly consistent with our empirical estimate for \snhope. This holds also assuming $\rm SFR_{10Myr}$ and $v_{\rm out}=\Delta v_{\rm off}+2\sigma$ for the calculation ($\dot{E}_{\rm out}\propto v_{\rm out}^3$, thus $5.6$ times higher than in the case of $v_{\rm out}=\Delta v_{\rm off}$). Similarly, the mass outflow rate is in agreement with our observational estimate ($\dot{M}_{\rm out}=0.2-5$ \myr, Table \ref{tab:results}). Taken with the findings described in the previous section, it seems that the levels of star formation over the last $\sim$10-100 Myr are sufficient to explain the properties of the outflows observed in \snhope\ (Fig. \ref{fig:sfh}). However, given the slow average outflow velocity ($\Delta v_{\rm off}\sim180$ \kms, lower than the stellar velocity dispersion ($\sigma_\star\sim250$ \kms), and the mass loading factor $\eta\lesssim1$, it is unlikely that the observed outflows are responsible for the quenching of star formation in \snhope.

\subsection{Extreme outflow properties might indicate undetected AGN activity in RUBIES-UDS-QG-z7}
The very large value of $\eta>10$ and $\dot{E}_{\rm out}\sim2\times10^{42}\,\mathrm{erg}\,\mathrm{s}^{-1}$ for \rubies\ give leeway to different interpretations. The mass outflow rate from \mgii\ ($\dot{M}_{\rm out}\sim269$ \myr\ for $v_{\rm out}=\Delta v_{\rm off}$, $1.8\times$ higher for a less conservative assumption on $v_{\rm out}$; Sect. \ref{sec:outflow_properties}) largely exceeds the value expected from SNe-driven winds ($\dot{M}_{\rm SN}=9^{+24}_{-8}$ and $<0.5$ \myr\ assuming $\rm SFR_{100Myr}$ and $\rm SFR_{10Myr}$, respectively), suggesting a contribution from other ejective mechanism, i.e., powered by an AGN. The injected energy is broadly consistent with $\dot{E}_{\rm SN}=2^{+6}_{-2}\times10^{43}\,\mathrm{erg}\,\mathrm{s}^{-1}$ considering a 10\% coupling with the ISM only by assuming $\rm SFR_{\rm 100Myr}$ and $v_{\rm out}=\Delta v_{\rm off}$ as representative, and the agreement is largely due to the loose constraints on the recent SFR estimate. In fact, adopting $v_{\rm out}=\Delta v_{\rm off}+2\sigma$ would increase $\dot{E}_{\rm out}$ by a factor of $\sim6$ and $\rm SFR_{10Myr}$ would set an upper limit on $\dot{E}_{\rm SN}<10^{42}\,\mathrm{erg}\,\mathrm{s}^{-1}$. The findings for \rubies\ are, thus,  more consistent with those obtained from mapping the neutral gas outflows using the sodium NaI D tracer in recently quenched galaxies at $z=2-3$,  carrying large amount of gas masses and energies away from galaxies \citep{belli_2024, davies_2024}, as simulations and models postulate, despite the absence of clear ongoing AGN activity.\\

\subsection{Caveats and future perspectives}
Combined with recent findings on outflows and galaxy quenching in the literature, our results for \snhope\ and \rubies\ contribute to describing this phenomenon as being as rich and complex as it is in the local Universe, despite the much shorter time frame in which it can occur. However, a few words of caution are due.\\ 

First, we stress that the measurements and simple calculations in the previous sections are affected by significant uncertainties and rely on several  assumptions. As an example, in the case of \snhope, the mass loading factor derived in this work significantly differs from the range presented in \citet[$\eta\sim7-30$]{wu_2025}, which drives us to different conclusions about the origin of the observed outflows. The disparity is mainly ascribable to the SFR estimate to compute $\eta$. Different assumptions in the modeling of the spectrophotometric data of \snhope\ lead to different SFR estimates than presented in \cite{wu_2025}. Moreover, as noted in Sect. \ref{sec:outflow_properties}, adopting SFRs averaged over 10 Myr or 100 Myr periods gives an idea of the expected systematic uncertainties on $\eta$ and the rest of the physical quantities depending on these parameters. While the time necessary for an outflow moving at $v_{\rm out} = \Delta v_{\rm off}$ to reach $R_{\rm out}$ would suggest to employ 10 Myr averaged SFRs, a timescale of 100 Myr is on the lower end of the lifespan of outflows in the galaxy halos as suggested by observational and theoretical studies \citep[][and references therein]{davis_2023}. The choice of $\rm SFR_{100Myr}$ also allows us to directly compare our findings with those in the literature compilation shown in Fig. \ref{fig:literature}. Together with the analysis of the origin of possible faint emission lines in the optical rest-frame spectrum and the mild lensing correction, this is where our analysis of \snhope\ mainly differs from that presented in \citet[see also our Appendix \ref{app:extended_snhope}]{wu_2025}.

Second, definitive proof of the effect of AGN  and stellar feedback on their hosts remains hard to constrain even with exquisite data and under consistent assumptions, given the widely different spatial and temporal scales of several competing processes involved. In particular, extreme outflow properties (e.g., velocities, masses, and energies) can be somewhat more confidently associated with the action of AGNs than to SNe, even in absence of typical signatures such as high optical line ratios or broad lines in their spectra, as in the case of \rubies. This is because of the short-term AGN variability (i.e., the imprint of an AGN on outflows can outlast its presence and detectability in recently quenched galaxies) and the fact that the feedback mode that quenches star-formation in galaxies can be characterized by low Eddington ratios in some models (\citealt{lagos_2025}, even if this is still matter of debate \citealt{choi_2017,delucia_2024,farcy_2025}). The variability and radiative inefficiency play also a role in the case of less extreme outflows conditions, such as those we reconstructed for \snhope\ and similarly found in normal star-forming galaxies. While in principle consistent with the energy, momentum, and mass provided by SN feedback, low values of these quantities do not exclude a significant contribution from AGNs a priori, especially for galaxy-scale outflows and for those models where gas entrainment in the circumgalactic medium plays a major role \citep{mitchell_2020}.\\

Promising approaches to countering the effect of the short-term AGN variability rely on proxies of the integrated, rather than instantaneous, effect of the growth of supermassive black holes (e.g., their masses; \citealt{bluck_2023}) on their host galaxies. Valid attempts tested in the local Universe \citep[e.g.,][]{wang_2024} are now extended to high-redshift samples \citep[see the discussions in][]{ito_2025_agn, baker_2025, onoue_2024}. Tying these properties with those of outflows in larger samples of galaxies similar to those presented here will be key to attenuate the effect of different spatial and temporal scales, and to obtain a more decisive evidence of how stellar and black hole masses grow and coevolve from star-forming to quiescent phases.\\

\section{Conclusions}

We have reported the detection of signatures of outflowing gas in the rest-UV and optical JWST/NIRSpec spectra of two massive ($M_\star \sim 10^{10.2}\,M_\odot$) and recently quenched galaxies: \snhope\ at $z=4.1061$ and \rubies\ at $z=7.2758$. New observations with the G235M grating, reduced with a customized version of the \textsc{msaexp} pipeline, allowed us to ascertain the redshift of \rubies, which had previously only been  constrained by low-resolution PRISM data.
The presence of gas outflows is revealed by blueshifted ionized magnesium, iron, and neutral sodium absorption features, which trace the neutral hydrogen phase. Our analysis of the spectra of these galaxies allowed us to reach the following conclusions.

\begin{itemize}
    \item Both targets show signatures typical of recently quenched star formation. The absence of strong emission lines and the reconstructed SFHs from spectrophotometric modeling predict low SFRs on short (10 Myr) and intermediate (100 Myr) scales ($\mathrm{SFR_{\rm 100Myr}\sim15}$ \myr). The quiescence of \snhope\ is further confirmed by the lack of strong dust continuum emission, which allows us to set an upper limit on the obscured SFR of $<12$ \myr. 
    \item The measured outflow velocities ($\sim180$ \kms) are consistent with those measured in post-starburst galaxies nearby ($z\lesssim 0.7$) and up to $z\sim1-3$ at fixed \mstar, SFR, and $\Sigma_{\rm SFR}$ without strong signatures of ongoing AGN activity. This suggests that outflows in galaxies quenched on short timescales are powered by similar mechanisms across cosmic time. This can be tested by tracing these scaling relations at $z\gtrsim2-3$ (in particular, the tightest one: $v_{\rm out}$-$\Sigma_{\rm SFR}$) with large spectroscopic samples of recently quenched systems.
    \item Under simple assumptions, we derive outflow masses, rates, loading factors, and energetics for both galaxies -- finding them to be rather different. For \snhope\ at $z=4.1$, benefitting from higher quality spectra, we find values that are in principle consistent with predictions for SN-powered outflows over timescales of 10-100 Myr. However, low values of the mass loading factor ($\eta\lesssim 1$) and velocities indicate that it is unlikely that the observed outflows are responsible for the star formation quenching. 
    \item For \rubies, the extreme mass loading factor ($\eta> 10$) suggests the contribution of an undetected AGN as a powering mechanism for the observed outflow. Interestingly, this is not reflected in an extremely high average outflow velocity. In fact, the observational estimate is consistent with that recorded for the outflow in \snhope\ and in local post-starburst galaxies with similar \mstar, SFR, and/or $\Sigma_{\rm SFR}$ values.
    \item Maybe more importantly, these case studies prove that it is possible to detect neutral outflowing gas in distant ($z>4$) post-starburst galaxies via Mg and Fe in the rest-frame UV, and Na in the optical, even with relatively inexpensive integrations of a handful of hours with medium resolution gratings with JWST/NIRSpec. The logical next step is to construct larger samples to test scaling relations as a function of cosmic time and compare outflow properties with model predictions for instantaneous and integrated SN and AGN  feedback; such work would mitigate the effect of the wide range of spatial and temporal scales involved in galaxy quenching.

\end{itemize}

\section*{Data availability}
The calibrated spectra and photometry are available via doi: \href{https://doi.org/10.5281/zenodo.15518189}{10.5281/zenodo.15518189}. The specific JWST and HST observations analyzed can be obtained from the Mikulski Archive for Space Telescopes at the Space Telescope Science Institute and accessed via doi: \href{https://archive.stsci.edu/doi/resolve/resolve.html?doi=10.17909/3ev1-rx30}{10.17909/3ev1-rx30}.

\begin{acknowledgements}
We thank the referee for their insightful comments that improved this article. FV warmly thanks Christy Tremonti for providing the measurements for the  literature sample shown in Fig. \ref{fig:literature}. FV is also indebted to Carlos G\'{o}mez-Guijarro, the co-principal investigator of the NOEMA program, for many years of stimulating and fruitful scientific discussions on the topics presented in this work -- and for his valued friendship. This work is based in part on observations made with the NASA/ESA/CSA James Webb Space Telescope. The data were obtained from the Mikulski Archive for Space Telescopes at the Space Telescope Science Institute, which is operated by the Association of Universities for Research in Astronomy, Inc., under NASA contract NAS 5-03127 for JWST. The observations are associated with programs \# 1176, 1837, 3567, 4233, and 4446. Support for program \# 3567 was provided by NASA through a grant from the Space Telescope Science Institute, which is operated by the Association of Universities for Research in Astronomy, Inc., under NASA contract NAS 5-03127. This work is also based on observations carried out under project number W23CU with the IRAM Interferometer NOEMA. IRAM is supported by INSU/CNRS (France), MPG (Germany) and IGN (Spain). We are grateful for the help received from IRAM staﬀ during observations and data reduction.
FV is grateful for the support of the Japanese Society for the Promotion of Science through the Fellowship (JSPS) S23108 and for the hospitality of Masayuki Tanaka and his group at the National Observatory of Japan, where part of this work has been conducted. 
Some of the data products presented herein were retrieved from the Dawn JWST Archive (DJA). DJA is an initiative of the Cosmic Dawn Center, which is funded by the Danish National Research Foundation under grant DNRF140. FV, KI, and PZ acknowledge support from the Independent Research Fund Denmark (DFF) under grant 3120-00043B. KEH acknowledges support from the Swiss State Secretariat for Education, Research and Innovation (SERI) under contract number MB22.00072.
SJ and GEM acknowledges the Villum Fonden research grants 37440 and 13160. MF and MH acknowledge funding from the Swiss National Science Foundation (SNF) via a PRIMA Grant PR00P2 193577 ``From cosmic dawn to high noon: the role of black holes for young galaxies''.
PFW acknowledges funding through the National Science and Technology Council grants 113-2112-M-002-027-MY2. This study was also supported by JSPS KAKENHI Grant Numbers JP22J00495, JP23K13141, and JP25K07361.
\end{acknowledgements}

\bibliographystyle{aa} 
\bibliography{bib_outflow_psb}

\begin{appendix}

\onecolumn
\section{Extended spectra of \snhope}
\label{app:extended_snhope}
For full transparency, we ran the customized \textsc{msaexp} pipeline also on the G140M and G235M spectra of \snhope. In this case, the advantage of extending the G140M spectrum is marginal compared with the gain obtained by processing the bluest spectrum of \rubies\ -- simply because a deep  coverage of the rest-frame optical wavelengths is already available. The G235M grating spectrum extension covers the \ha\ Balmer line and the \nii$\lambda\lambda$6549,6585 doublet. However, in the case of \snhope, the extension comes with an uncertain absolute flux calibration and a systematic color gradient compared with the PRISM spectrum. For reference, we reran the same pPXF modeling described in Sect. \ref{sec:modeling}  and estimate a stellar velocity dispersion of $\sigma_\star=278\pm23$ \kms, 10\% higher than, but in agreement with the value in Table \ref{tab:results}. We also estimate  observed (thus, magnified) line fluxes of \neiii$\,=2.8^{+2.6}_{-2.4}$, \oii$\,=4.0^{+3.0}_{-2.8}$, \hb$\,=2.3^{+2.2}_{-2.3}$, \oiii=$\,8.9^{+2.6}_{-2.8}$, \ha$\,=15.4^{+3.2}_{-3.5}$, and \nii$\lambda6585\,=37.5^{+5.0}_{-4.9}$ all expressed in units of $10^{-19}\,\rm{erg\,cm^{-2}\,s^{-1}}$. We stress that the flux calibration uncertainties, and especially the color term, influence the line fluxes and their ratios non trivially, as the stellar continuum absorption features have to be modeled in order to derive the line emission. This explains the difference in the \hb\ flux measurement and its ratio with \oiii\ ($\mathrm{log}($\oiii/\hb$)>-0.37$ ratio at $3\sigma$ in the extended spectrum). At face value, we estimate $\mathrm{log([NII]/H\alpha)}=0.39^{+0.13}_{-0.11}$ (where the uncertainties are purely statistical), which would place \snhope\ beyond the extrapolation to $z=4$ of the line dividing star-forming galaxies and AGNs in the BPT diagram \citep{baldwin_1981} presented in \cite{kewley_2013p}. We note that the parameterization in \cite{kewley_2013p} is not calibrated against data at $z=4$ and should be taken with a grain of salt. If confirmed, these emission line ratios would be potential evidence of a faint AGN-powered emission, or at least non-star-formation-powered ionized emission, as first proposed in \cite{wu_2025}. Future, deeper observations will test this possibility.
\begin{figure*}[h!]
    \includegraphics[width=\textwidth]{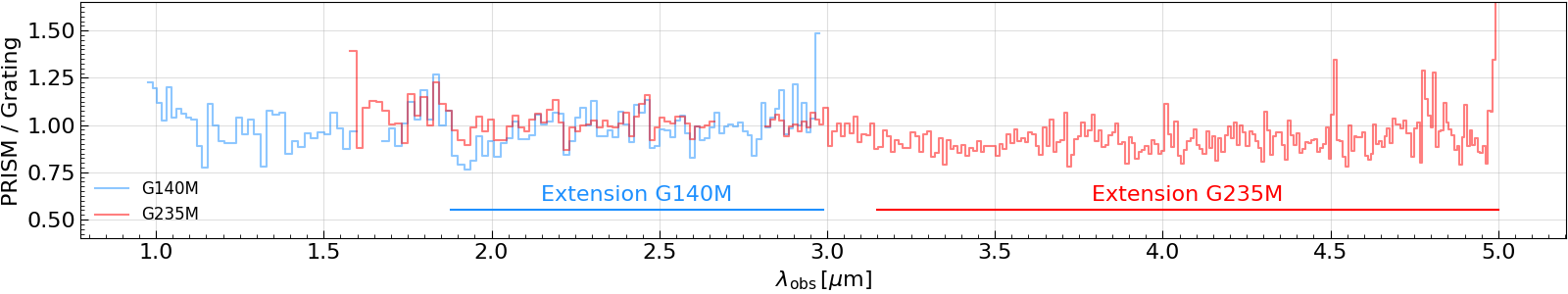}
    \includegraphics[width=\textwidth]{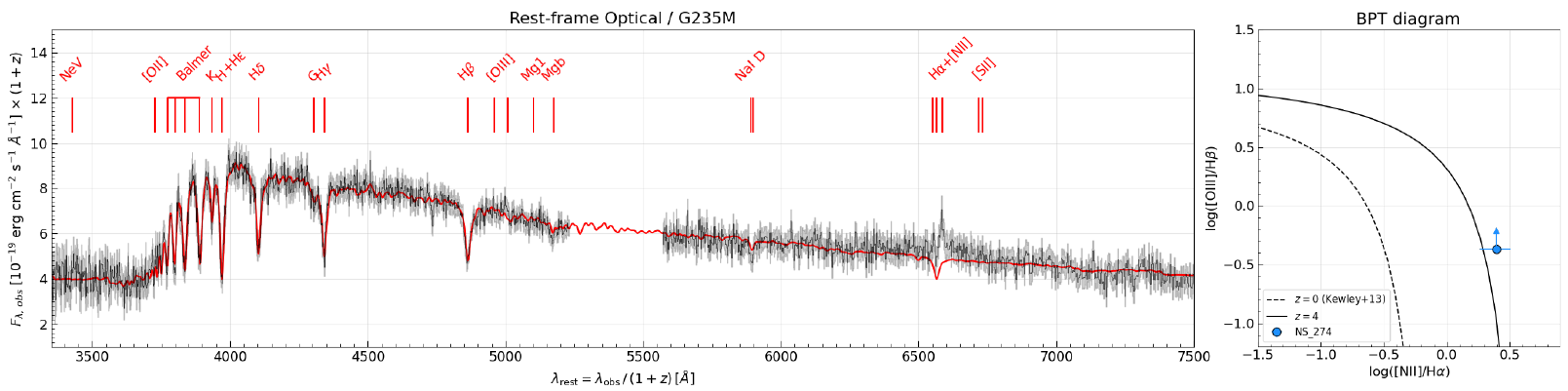}
    \caption{\text{Top:} Flux density ratio between the PRISM and the G140M (blue line) and G235M (red line) grating spectra in the custom reduction of \snhope. We mark with a solid line the extended wavelength coverage. {Bottom:} Extended G235M spectrum of \snhope\ (black) and its uncertainties (gray). The red line indicates the best-fit pPXF model of the stellar continuum. The panel on the right shows the location of the source in the BPT diagram based on the measurements in the extended G235M spectrum. The dashed  and solid black lines indicate the empirical parameterization of the line dividing star-forming galaxies and AGNs by \cite{kewley_2013p} at $z=0$ and $z=4$, respectively.}
    \label{app:fig:extended_snhope}
\end{figure*}

\begin{figure*}
    \centering
    \includegraphics[width=0.48\textwidth]{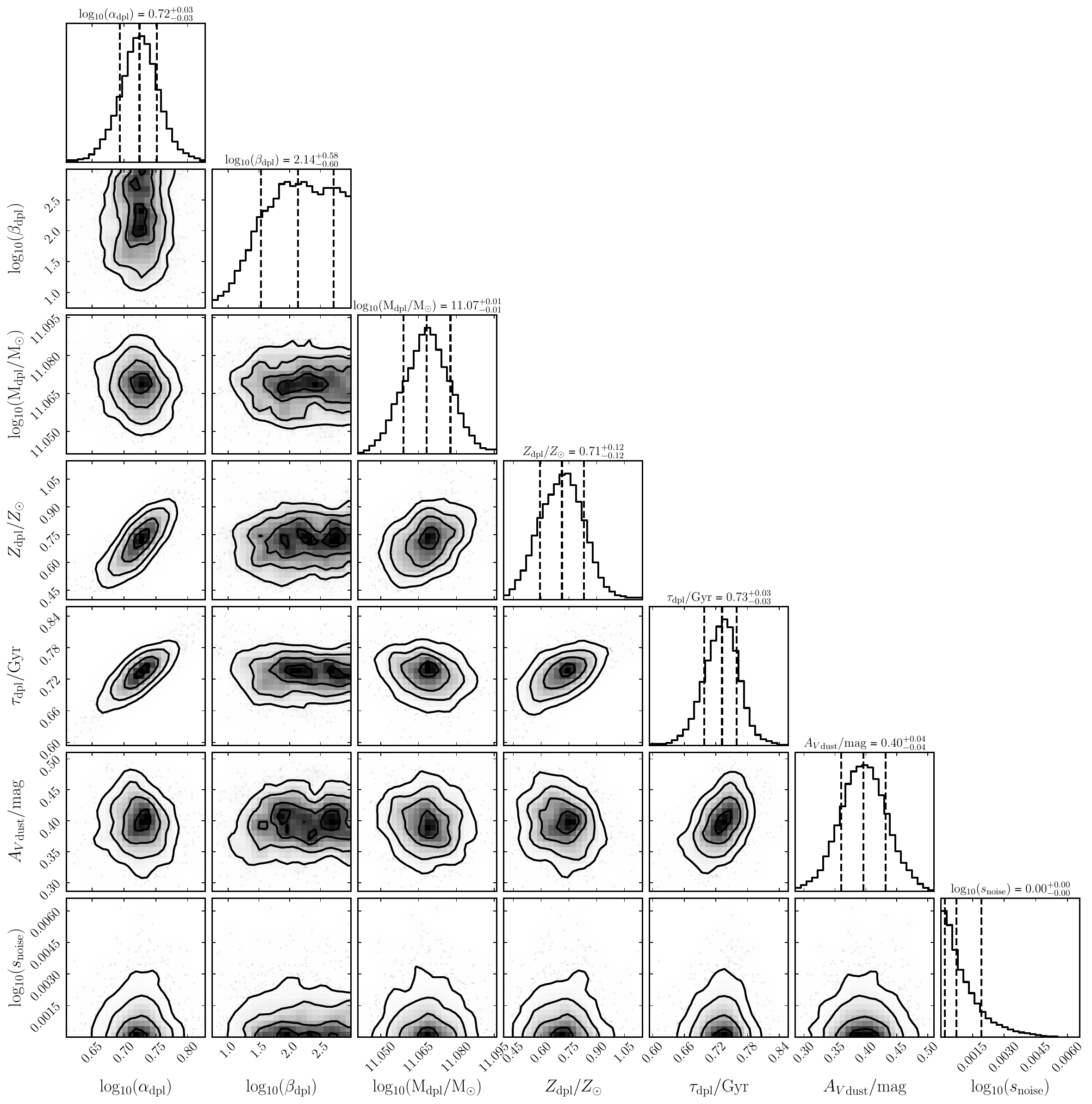}
    \includegraphics[width=0.48\textwidth]{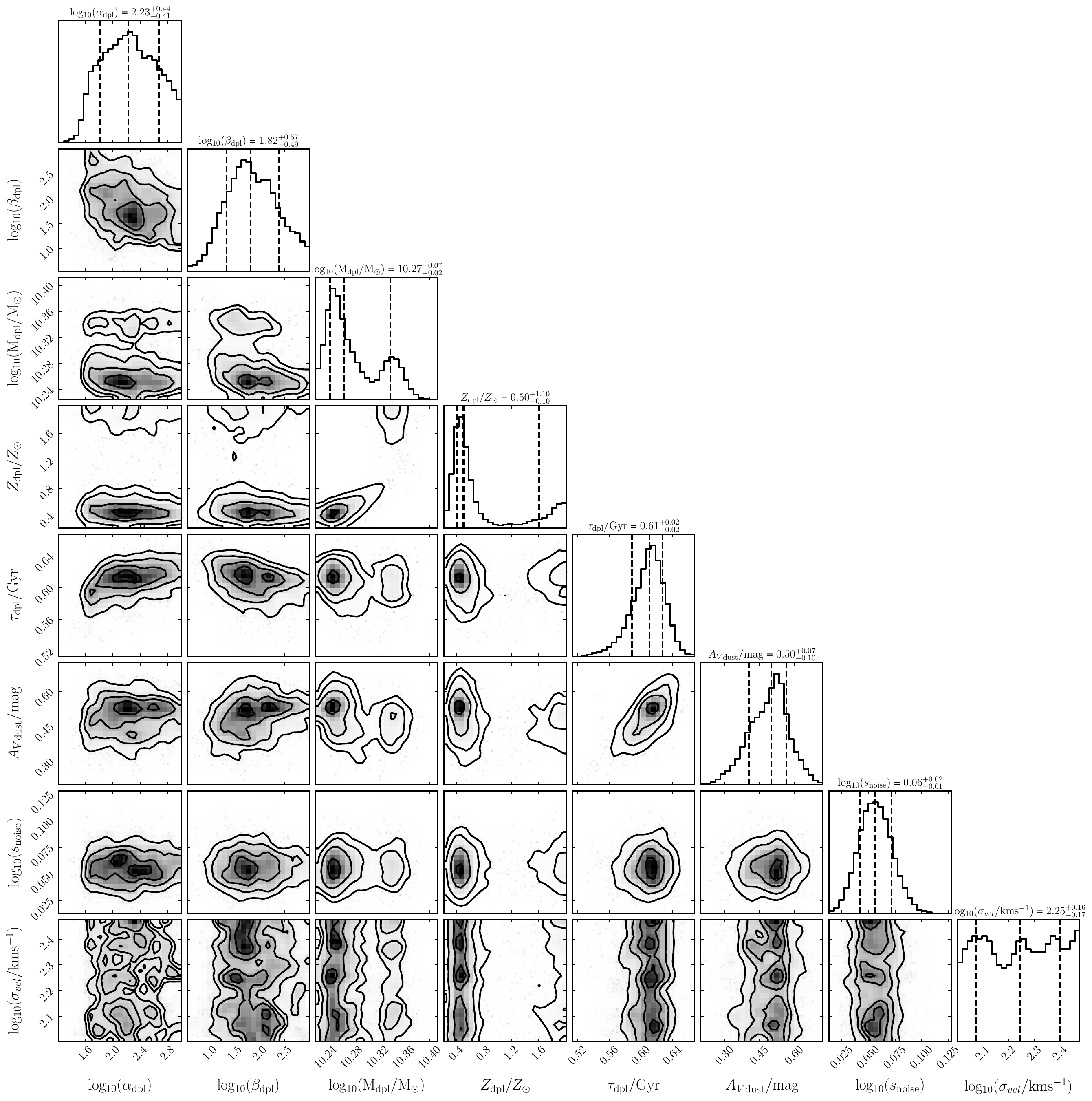}
    \caption{Corner plots showing the posterior distributions of the parameters constrained with \textsc{Bagpipes} as detailed in Sect. \ref{sec:modeling}. Left: \snhope\ at $z=4.1061$; Right: \rubies\ at $z=7.2758$.}
    \label{app:fig:corner}
\end{figure*}

\begin{figure*}[h!]
    \begin{minipage}[]{0.48\textwidth}
    \centering
    \includegraphics[width=\columnwidth]{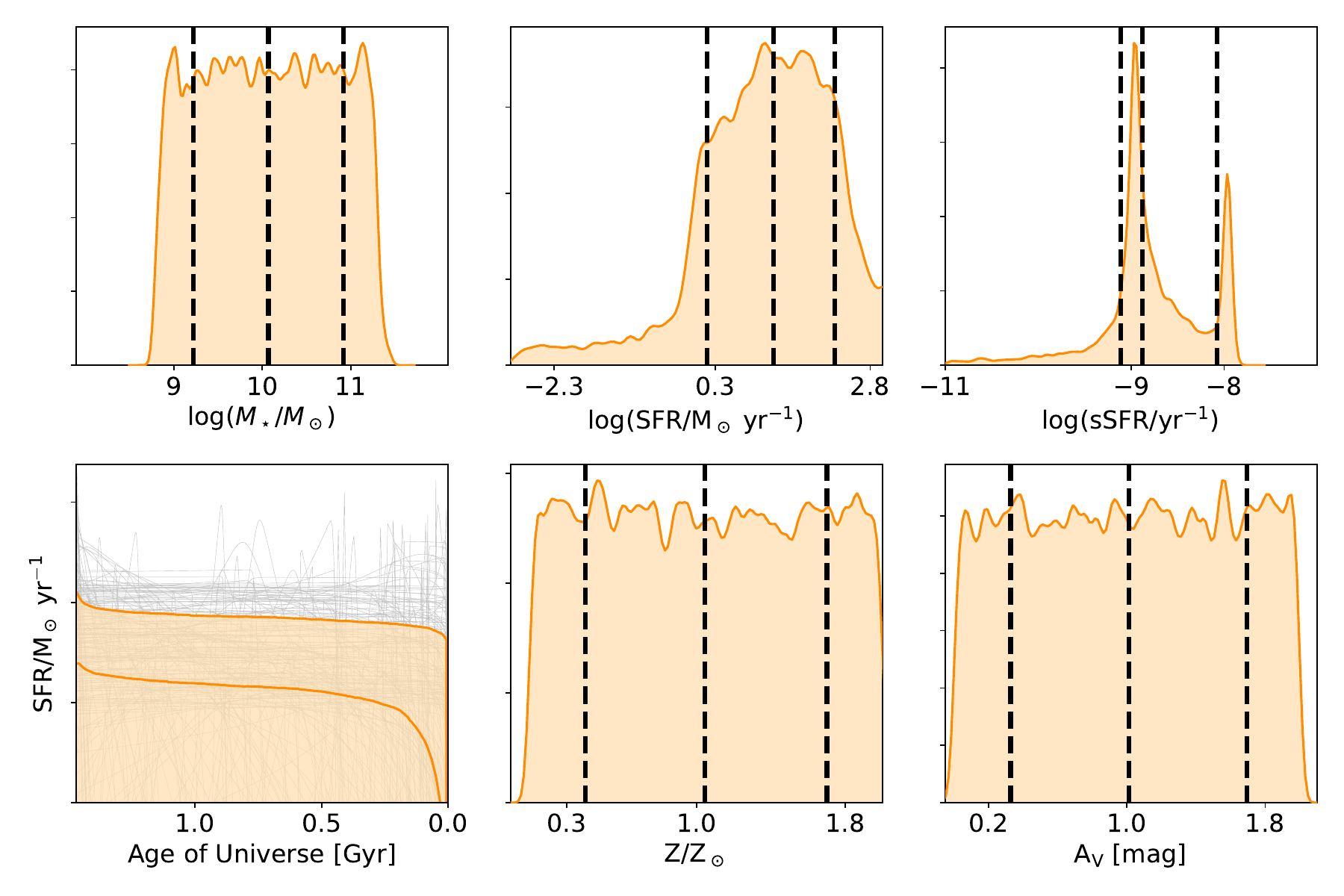}
    \caption{Priors on the parameters listed in Table \ref{tab:priors} for \snhope. Dashed black lines indicate the 16, 50, and 84\% percentiles of the distributions. In the SFH panel, we show randomly extracted SFHs in gray. The orange area shows the 16-84\% inter-percentile range. Similar priors were applied to model the spectrum and photometry of \rubies.}
    \label{app:fig:priors}
    \end{minipage}
    \begin{minipage}[]{0.48\textwidth}
    \centering
    \includegraphics[width=\columnwidth]{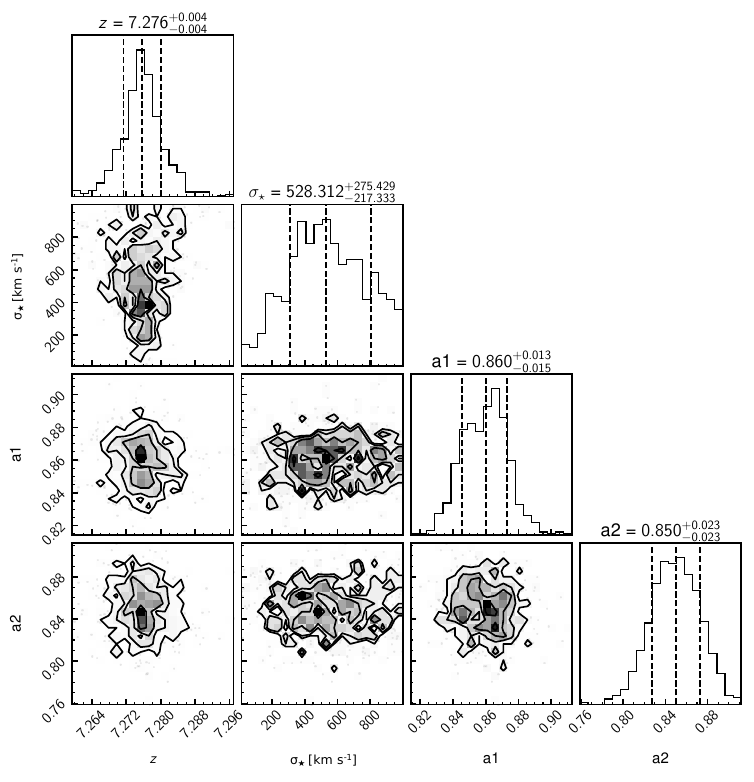}
    \caption{Corner plot showing the posterior distributions of the free parameters of the template scaling used to determine the redshift of \rubies.}
    \label{app:fig:zspec_corner}
    \end{minipage}
\end{figure*}

\begin{figure*}
    \centering
    \includegraphics[width=\textwidth]{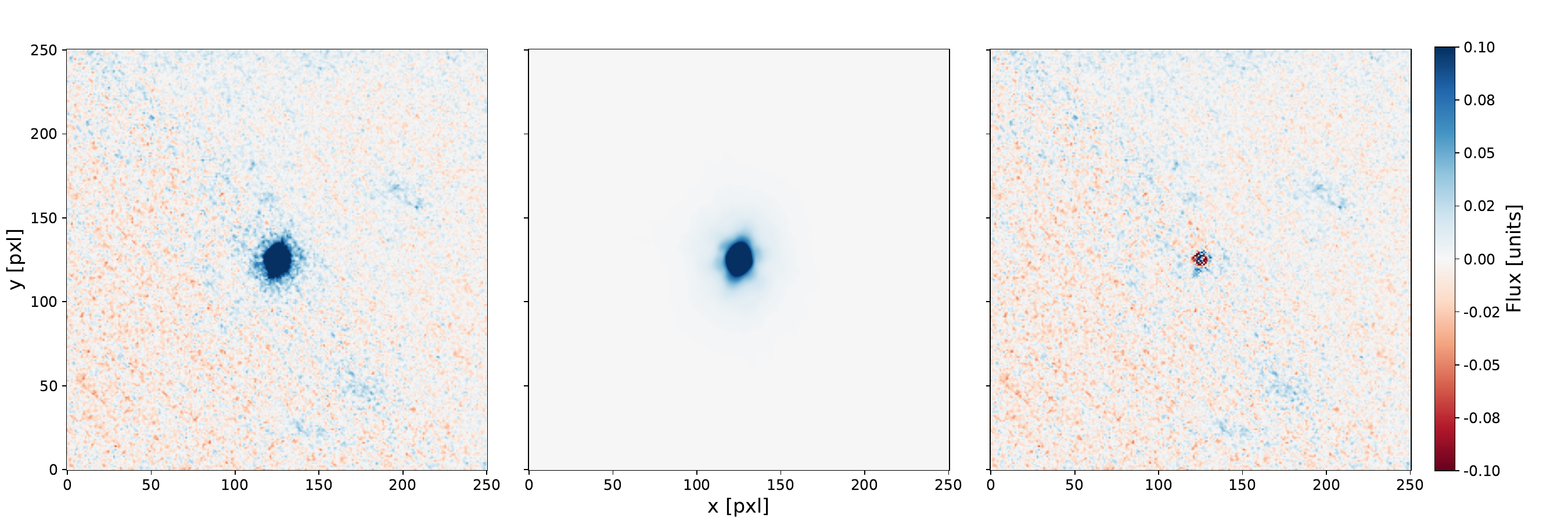}
    \caption{S\'{e}rsic modeling of \snhope. \textit{Left:} $5"\times5"$ ($\sim35\times35$ kpc at $z=4.1$) F200W cutout image of our target. \textit{Center:} Best-fit S\'{e}rsic model (Sect. \ref{sec:structure}). \textit{Right:} Residuals computed by subtracting the best-fit model from the input image.}
    \label{app:fig:pysersic}
\end{figure*}

\noindent
\begin{minipage}[t]{0.48\textwidth}

\section{Modeling supplementary material}
\label{app:bestfit_modeling}

In Fig. \ref{app:fig:corner} we show the posterior distributions of the parameters obtained by modeling the spectra and photometry of \snhope\ and \rubies\ as detailed in Sect. \ref{sec:modeling}. Properties of the priors are reported in Table \ref{tab:priors}. We show them in Fig. \ref{app:fig:priors}. 
In Fig. \ref{app:fig:zspec_corner} we report the results of the joint best-fit modeling of the G235M and G395M spectra of \rubies\ to determine its exact spectroscopic redshift (see Sect. \ref{sec:spectroscopy}).
Finally, Fig. \ref{app:fig:pysersic} shows the NIRCam/F200W image of \snhope, the best-fit S\'{e}rsic described in Sect. \ref{sec:structure}, and the residuals.
\end{minipage}
\hfill
\noindent
\begin{minipage}[t]{0.48\textwidth}
    \centering
    \captionof{table}{Free parameters and priors for the spectrophotometric modeling.}
    \label{tab:priors}
    \begin{tabular}{lcc}
    \toprule
    \toprule
        Free parameter & Prior & Limits \\
    \midrule
        $\mathrm{log}(M_{\rm formed}/M_\odot)$ & Uniform & (9, 11.5)\\
        $A_{\rm V}$ &                            Uniform & (0,  2)\\
        $Z / Z_\odot$ &                          Uniform & (0.1,  2)\\
        $\tau / \mathrm{Gyr}$ &                  Uniform & (0,  $t(z_{\rm obs})$)\\
        $\alpha$, $\beta$ &                  Logarithmic & (10$^{-2}$, 10$^{-3}$)\\
        Noise rescale $s_{\rm noise}$ &                  Logarithmic & (1, 10)\\
    \bottomrule
    \end{tabular}
\end{minipage}

\FloatBarrier
\twocolumn

\end{appendix}

\end{document}